\documentclass[pdflatex,sn-mathphys-num]{sn-jnl}

\usepackage{graphicx}%
\usepackage{multirow}%
\usepackage{amsmath,amssymb,amsfonts}%
\usepackage{amsthm}%
\usepackage{mathrsfs}%
\usepackage[title]{appendix}%
\usepackage{xcolor}%
\usepackage{textcomp}%
\usepackage{manyfoot}%
\usepackage{booktabs}%
\usepackage{algorithm}%
\usepackage{algorithmicx}%
\usepackage{algpseudocode}%
\usepackage{listings}%
\usepackage{subcaption}
\usepackage[table]{xcolor}
\usepackage{colortbl}

\definecolor{darkred}{RGB}{139, 0, 0}
\definecolor{lightyellow}{RGB}{255, 255, 200}
\definecolor{lightblue}{RGB}{200, 220, 255}
\definecolor{darkblue}{RGB}{0, 0, 139}
\usepackage{tikz}
\usetikzlibrary{shapes.geometric, arrows.meta, positioning, fit}
\tikzstyle{block} = [rectangle, rounded corners, minimum width=3cm, minimum height=1.2cm, text centered, draw=black, fill=blue!10]
\tikzstyle{decision} = [diamond, minimum width=2.5cm, minimum height=1.2cm, text centered, draw=black, fill=orange!20]
\tikzstyle{arrow} = [thick,->,>=Stealth]

\begin{document}

\title[Article Title]{Oscillatory Hierarchical Reservoirs for Human-like Rhythm Perception and Anticipation}


\author*[1]{\fnm{Zhongju} \sur{Yuan}}\email{zhongju.yuan@ugent.be}

\author[2,3]{\fnm{Geraint} \sur{Wiggins}}\email{geraint.wiggins@vub.be}

\author[1]{\fnm{Dick} \sur{Botteldooren}}\email{dick.botteldooren@ugent.be}

\affil*[1]{\orgdiv{WAVES Research Group}, \orgname{Ghent University}, \orgaddress{ \city{Gent}, \postcode{9000}, \country{Belgium}}}

\affil[2]{\orgdiv{AI Lab}, \orgname{Vrije Universiteit Brussel}, \orgaddress{\city{Brussel}, \postcode{1050}, \country{Belgium}}}

\affil[3]{\orgdiv{EECS}, \orgname{Queen Mary University of London}, \orgaddress{\city{London}, \country{UK}}}


\abstract{Rhythm is a fundamental aspect of human behaviour, present from infancy and deeply embedded in cultural practices. Rhythm anticipation often occurs before surface event onsets, yet most neuroscience and artificial intelligence studies focus on metronome-based tasks, with less attention to complex musical rhythms. To address this gap, we propose a hierarchical oscillator-based computational model for selected aspects of complex rhythm perception. The model uses coupled neurons that generate oscillations across four layers corresponding to Tatum, Tactus, Motor, and Higher Cognition levels. We evaluate the model using representative rhythm patterns spanning different tempi and perceptual ranges. The model maintains stable entrainment and exhibits selected rhythm-relevant behaviours, including anticipatory movement-timing predictions, timing deviations under tactus-surface conflict, and suppression of movement-timing predictions at anticipated-but-silent tatum-level pulses. In addition, Motor Layer beta-band activity shows rhythm- and tempo-dependent modulation, which we report as a qualitative exploratory comparison with human beta activity.}

\maketitle

\section*{Introduction}

When a child sings a nursery rhyme or a musician performs Mozart, the neural mechanisms underlying the perception and production of music pose intriguing questions for cognitive neuroscience. The interaction between auditory and motor systems has garnered significant attention. For instance, playing a musical instrument, such as the drums, requires precise coordination between the auditory signal (such as a musical note or rhythmic event) and the motor action (the movement of the arm to strike the drum pad). Three fundamental motor control functions are essential for performing rhythmic patterns: timing, sequencing, and movement organization. Accurate timing governs the rhythmic structure, while sequencing and organization are crucial for executing each note in the rhythmic sequence.

These motor control functions correspond to different levels of metric rhythmic pattern perception. Metric rhythmic patterns are typically divided into three hierarchical levels: the \textit{Tatum}, representing the smallest isochronous subdivision; the \textit{Tactus}, corresponding to the perceived pulse, or tactus, that listeners naturally entrain to; and the \textit{Measure}, the highest metric level that organizes tactus pulses into recurring groupings and is most directly linked to motor planning and movement organization~\cite{london2012hearing, vuust2014rhythmic}. We distinguish surface event onsets in the binary rhythm pattern from the perceived beat or tactus. We use ``tatum'' for the smallest time unit or target tatum interval, ``tactus'' for the perceived pulse, and ``beat'' only for the tactus or perceived pulse unless explicitly referring to terminology used in the human rhythm literature. We use ``surface event onset'' for a tatum-aligned position carrying a \texttt{1} in the binary pattern encoding, and ``silent tatum position'' for a tatum-aligned \texttt{0}. We use ``tatum-level pulse'' for a possible pulse-like event or expectation at the tatum scale. An ``anticipated-but-silent tatum-level pulse'' denotes a tatum-aligned position at which the ongoing grouping structure creates an expectation for a pulse-like event, although the actual binary pattern contains no surface event. Tactus pulses are often structured into patterns that create hierarchical periodicities, such as march-like patterns with strong tactus pulses every two pulses or waltz-like patterns with strong tactus pulses every three pulses~\cite{patel2014evolutionary, bouwer2016disentangling, celma2016look}. The Meter, rooted in the perception and production of the pulse or tactus, is flexible across a range of frequencies~\cite{patel2014evolutionary}, with the strongest rhythmic responses occurring within 400–1200 ms intervals. Across cultures, music unfolds within a higher-order metric structure, creating musical expectations that enable individuals to predict and organize future motor behaviours~\cite{large2002perceiving, breska2020context, proksch2020motor}. Human participants demonstrate rhythmic synchronization with these pulses via precise temporal predictions, typically within tens of milliseconds~\cite{van2013adaptation, palmer2022we}.

Rhythm perception and production involve several brain regions in a hierarchical manner~\cite{london2012hearing, fontolan2014contribution}. Neuropsychological and neuroimaging studies have revealed distinct pathways originating in the primary auditory cortex and projecting to various targets~\cite{rauschecker2000mechanisms, ito2022spontaneous}. Rhythm processing recruits auditory--motor circuits involving the basal ganglia and cerebellum~\cite{grahn2007rhythm}. The basal ganglia and supplementary motor area (SMA) play critical roles in interval timing~\cite{buhusi2005makes, lewis2003distinct}, and the putamen shows particularly strong engagement during tactus-based rhythm processing, with differential activation for metric versus non-metric rhythms~\cite{grahn2007rhythm, grahn2009feeling}. A cortico-striatal network involving the putamen, SMA, and premotor cortex (PMC) is thought to support temporal prediction and tactus detection~\cite{grahn2013finding, teki2012unified}, and this engagement of motor regions occurs even in the absence of overt movement~\cite{grahn2007rhythm, geiser2012corticostriatal, kung2013interacting, matthews2020sensation}. Crucially, top-down knowledge of internalized metric structure selectively enhances beta-band responses at imagined tatum-level pulse positions even when the physical stimulus is invariant~\cite{iversen2009top}, implicating beta oscillations as a bidirectional interface between endogenous metrical expectations and exogenous auditory processing. Both the pre-SMA and SMA are implicated in organizing complex movement sequences~\cite{sakai2004emergence, kennerley2004organization, ohbayashi2021roles}, while the premotor cortex is associated with motor prediction~\cite{janata2003swinging, schubotz2003functional, zimnik2021independent}.

Studies investigating rhythms with syncopation, metric ambiguity, and polyrhythmic structures report engagement of additional cortical and subcortical regions~\cite{chen2008moving, morillon2014motor, braun2022rhythm, large2023dynamic}, suggesting that these structures support hierarchical processing and predictive mechanisms. Complex rhythms, especially those with non-integer ratios, challenge auditory-motor synchronization, leading to greater variability and reduced synchronization~\cite{mathias2020rhythm}. Moreover, beta-band (15–30 Hz) activity has emerged as a biomarker of temporal predictions and omissions in auditory rhythm processing~\cite{doelling2015cortical, merchant2015finding, fujioka2015beta}.

Beyond descriptive neural correlates, computational models have sought to explain the mechanisms underlying rhythm perception. The Gradient Frequency Neural Network (GrFNN) framework~\cite{large2015neural, large2023dynamic} employs banks of nonlinear Hopf oscillators tuned to a gradient of frequencies, enabling hierarchical entrainment to complex rhythmic structures and successfully accounting for syncopation perception, meter induction, and cross-cultural rhythm processing. Underpinning GrFNN, Neural Resonance Theory~\cite{large2010neurodynamics} posits that rhythm perception emerges from resonant interactions between intrinsic neural oscillations and external periodic stimuli, validated through neural entrainment studies showing phase-locking of cortical oscillations to musical beats~\cite{nozaradan2012selective}. Other approaches include oscillatory timing frameworks~\cite{van1999resonance} and predictive-coding accounts in which listeners infer underlying rhythmic structure by continually updating expectations about event timing in response to prediction errors~\cite{vuust2014rhythmic, cannon2021expectancy}. However, it remains unclear whether a single generic oscillatory substrate, adapted only through rhythm-specific readout calibration, can jointly support hierarchical entrainment, anticipatory event timing, and suppression of anticipated movement-timing predictions at predicted silent positions.

To address this gap, we propose a physics-inspired, oscillation-driven framework that models hierarchical rhythm perception through coupled dynamical systems. The structure is shown as Fig.~\ref{fig:2}. The system is based on an echo state network that supports both oscillatory entrainment and inhibition. The physics-inspired design is motivated by two key requirements that existing models leave unmet. First, a generic oscillatory substrate is needed that naturally spans the full range of metric timescales without hand-tuning: the spatial gradient in wave speed across the Finite-Difference Time-Domain (FDTD) grid provides this, instantiating a continuous family of resonant modes from tatum- to measure-level frequencies through the physics of wave propagation. Second, the architecture must not only entrain to perceived pulses but also suppress an anticipated movement-timing prediction at tatum-level positions where no surface sound event occurs. In the model, the Higher Cognition Layer predicts these silent positions and provides a top-down control signal that suppresses the corresponding movement-timing prediction generated by the Motor Layer. This predictive inhibition capability is not included in the baseline GRU and Transformer comparisons used here. Unlike these purely perceptual baselines, our model modulates its internal eigenspectrum to achieve both rhythmic anticipation and suppression at anticipated-but-silent tatum-level pulses. This inhibition mechanism is mathematically grounded in the dynamic tuning of damping coefficients in the network's state-space formulation and is discussed as a computational mechanism, with beta-band activity used as a qualitative biological reference point.

Specifically, our model consists of four hierarchical layers (Fig.~\ref{fig:3}) that reflect distinct levels of rhythmic analysis—from smallest perception time unit (Tatum) to long-term metric expectations (Higher Cognition). In the Results section, we first provide a reader-oriented overview of how these layers interact and why each is needed before turning to quantitative evaluation, while the full mathematical specification is reserved for the Materials and Methods section. The resulting architecture learns to generate temporally aligned movement-timing predictions and reproduces selected qualitative features of timing behavior in complex, syncopated patterns.

For each new rhythm, calibration is restricted to the layer-specific readouts while the recurrent reservoir and input weights remain fixed. The Tatum readout is first pretrained on synthetic pulse trains and is then calibrated to the new rhythm for 6 seconds using a tatum-level target in a 200~ms-ahead anticipatory reference frame. The Tactus readout is calibrated to a tactus-like pulse target derived algorithmically from the symbolic rhythm. The Motor readout is calibrated to predict the rhythm's surface-event onsets 200~ms ahead, whereas the Higher Cognition readout is calibrated to reproduce the binary event--silence sequence defined by the symbolic rhythm. After calibration, all readouts are held fixed during evaluation.

The central idea is that the physics-inspired reservoir provides a structured family of resonant modes, while layer-specific online readout calibration selectively exploits the resonances most relevant to the rhythm being processed. We hypothesize that such processing involves spatially distributed wave-like dynamics, with different neural populations resonating at frequencies corresponding to distinct metric levels, consistent with Neural Resonance Theory~\cite{large2015neural} and hierarchical auditory-motor coupling~\cite{london2012hearing}. Within this framework, anticipated-but-silent tatum-level pulses are implemented as suppression in the Motor Layer. We therefore interpret this mechanism as a computational form of top-down suppression, while treating Motor Layer beta-band activity as a qualitative comparison with rhythm-related sensorimotor oscillations.

\section*{Results}
\subsection*{Model overview and processing logic}
The goal of the proposed architecture is to transform a repeating surface rhythm into a hierarchy of internal temporal representations that culminate in a movement-timing prediction. The Motor Layer prediction is therefore a model-generated timing signal. As shown in Fig.~\ref{fig:2}, all four layers share the same physics-inspired reservoir: a Finite-Difference Time-Domain (FDTD)-structured echo state network composed of $p$ neurons and intermediate $o$ neurons arranged on a spatial grid. The gradient in wave speed across this grid creates a structured bank of resonant timescales, allowing the reservoir to express dynamics ranging from fast tatum-like responses to slower meter-level responses without hand-tuning separate oscillators. Fig.~\ref{fig:3} then shows how this common dynamical substrate is used differently across the four layers through distinct readouts and inter-layer connections. Full equations and parameter settings are provided in the Methods and Materials section; here we focus on the processing logic needed to interpret the results.

Each layer contributes a distinct role in the proposed hierarchy. The Tatum Layer provides a fine-grained temporal scaffold, while the Tactus Layer organizes this scaffold into a slower metric grouping. The Higher Cognition Layer represents event–silence expectations over the learned pattern structure, and the Motor Layer converts these hierarchical timing signals into movement-timing predictions. This separation is important because rhythm processing cannot be reduced to local event tracking alone: tactus-level grouping can create internal timing expectations that interact with, and sometimes compete with, the surface rhythm. Therefore, Motor Layer predictions that do not coincide exactly with surface event positions are interpreted in relation to the model’s internal metric structure. In this framework, the Tatum Layer supplies temporal resolution, the Tactus Layer supplies metric organization, the Higher Cognition Layer supplies predictive context, and the Motor Layer expresses their combined influence on predicted movement timing.

The processing scheme for a previously unseen rhythm is intentionally minimal: the recurrent reservoir weights $W$ and input weights $W_{\mathrm{in}}$ remain fixed, while only the layer-specific readouts are adapted. This separates the general oscillatory substrate from the rhythm-specific calibration required to interpret a particular pattern. After calibration, the readouts are held fixed and the model is evaluated in inference mode. The Motor Layer output is defined as a movement-timing prediction. The Results therefore focus on how this calibrated hierarchy supports tatum detection, tactus-like grouping, predicted movement timing, and suppression of movement-timing predictions at predicted silent positions.

The results are organized to follow this processing hierarchy. We first test whether the Tatum Layer generalizes to unseen rhythms, then examine how closely the calibrated Tactus Layer reproduces the algorithmically derived grouping imposed through its calibration target, next evaluate the anticipatory movement-timing predictions generated by the Motor Layer, and finally analyze Higher Cognition-guided suppression of movement-timing predictions and Motor Layer beta dynamics at anticipated-but-silent tatum-level pulses.

To evaluate the performance of the proposed model on complex rhythmic patterns, we selected a diverse set of binary-encoded sequences spanning different lengths (6--12 tatums) and varying in metric regularity: some with clear binary grouping (\texttt{101001}, \texttt{10101001}), others with syncopation (\texttt{101000}, \texttt{10110010}), and one with metric ambiguity (\texttt{10010}). These sequences were based on an isochronous rhythm whose smallest unit served as the tatum, and the model was evaluated on seven representative patterns: three simpler ones (\texttt{101001}, \texttt{101000}, and \texttt{10101001}), three more challenging ones (\texttt{1001001010}, ``Mission Impossible''; \texttt{10110010}, ``Normal Rock''; and \texttt{101010100100}, ``Normal Jaz''), and the irregular pattern \texttt{10010}. Model performance was tested at 60, 100, and 140 BPM, corresponding to tatum intervals of 500, 300, and 214~ms, respectively, which together span a representative human beat-perception range~\cite{patel2014evolutionary}.

\subsection*{Evaluation metrics}

To evaluate the movement-timing predictions generated by the model, three key metrics are employed, as detailed in this section. Resultant Vector Length and Mean Asynchrony are used to assess the model's performance for predicted movement onsets corresponding to event onsets in the given complex patterns. Across layers, inter-beat interval (IBI) denotes the interval between successive detected output pulses or reference onsets, whereas inter-beat deviation (IBD) denotes the deviation between successive model-generated intervals and the corresponding reference intervals. This layer-dependent definition enables consistent comparison across Tatum-, Tactus-, and Motor-level outputs, whose reference sequences differ in temporal scale and functional role.

Synchronization was quantified by comparing model-generated output onsets with a layer-specific operational reference sequence. For the Tactus Layer, this reference corresponds to the empirically computed tactus-like grouping target; for the Motor Layer, it corresponds to the target event onsets in the rhythmic pattern.

\subsubsection*{Normalized Timing Deviation}
We denote the reference onset sequence by $B$ and the corresponding matched model-generated onset sequence by $M$. For each reference onset, the closest model-generated onset is selected to form the matched sequence. The normalized timing deviation is then computed as the relative timing difference between each matched output onset and its corresponding reference onset.
Since the model may miss target onsets or introduce extra output onsets compared to the mathematically strict reference sequence, many standard metrics fail to provide an accurate evaluation. To address this, for each reference onset $B_n$ in $B$, the closest model-generated onset is identified, resulting in a matched onset list $M$. Additionally, the index of each matched onset in the reference sequence is collected in an index list $I$. This approach ensures that synchronization is evaluated based on the closest corresponding onsets, allowing for a more robust and meaningful assessment.

For all reference onsets in $B$, the relative phase angles are calculated using Eq.~\eqref{eq:relative_phase_angle}\cite{heggli2019kuramoto, rosso2021neural}:
\begin{equation} 
\phi_n = 360 \times \frac{M_{n} - B_{n}}{B_{n+1} - B_{n}}, \label{eq:relative_phase_angle} 
\end{equation}
where $B_n$ denotes the $n$-th onset in the reference sequence and $M_n$ denotes the corresponding matched model-generated onset. The reference sequence defines the normalization interval $B_{n+1}-B_n$. For the Tactus Layer, this interval corresponds to the algorithmically derived grouping period; for the Motor Layer, it corresponds to the interval between successive target event onsets in the rhythmic pattern. A negative phase angle indicates that the model's generated onset occurs ahead of the corresponding reference onset, whereas a positive phase angle suggests that the model's onset lags behind the target timing. In the Motor Layer, this quantity is interpreted as a normalized event-level timing deviation.

\subsubsection*{Resultant Vector Length} The resultant vector length quantifies the stability of relative phase angles over time. A unimodal distribution corresponds to a high resultant vector length, while uniform or bipolar distributions result in a low value~\cite{berens2009circstat}. The resultant vector length, $R$, is calculated using the relative phase angles as shown in Eq.~\eqref{eq:resultant_vector_length}:
\begin{equation} 
R = \left| \frac{1}{N} \sum_{n=1}^{N} e^{i \phi_n} \right|, \label{eq:resultant_vector_length} 
\end{equation}
where $N$ is the number of target event onsets in the given rhythm signal. The value of $R$ ranges from 0 to 1, with 1 indicating perfect synchronization over time and 0 representing the absence of synchronization.

\subsubsection*{Mean Asynchrony} Mean asynchrony measures the average timing difference between the model-generated onsets and the corresponding onsets of the given complex rhythm, expressed in milliseconds. It is computed as follows:
\begin{equation} 
\text{Mean Asynchrony} = \frac{1}{N} \sum_{n=1}^{N} (M_n - B_n), \label{eq:mean_asynchrony} 
\end{equation}
where $B_n$ denotes the $n$-th onset in the reference sequence and $M_n$ denotes the corresponding matched model-generated onset. A negative value indicates an anticipatory response.

\subsubsection*{Interval Deviation} Although the abbreviation IBD is retained in the figures, this metric is used operationally as an inter-beat interval deviation measure and does not imply that all evaluated onsets are tactus-level beats. IBD quantifies how strongly the spacing of successive model-generated onsets deviates from that of the corresponding reference sequence. Its sensitivity to extra or missing output onsets makes it a measure of interval-level mismatch~\cite{rosso2021neural}.

To calculate IBD, the timing of matched reference onsets in $B$ is used together with the corresponding matched model-generated onsets in $M$. If the model generates an extra onset between two matched outputs, the interval between successive output onsets will deviate from the corresponding reference interval, increasing IBD even when the broader pulse level may remain plausible. This makes IBD an interval-deviation measure that can be applied consistently across layers. The IBD is calculated as follows:
\begin{equation} 
\text{Inter-Beat Deviation (IBD)} = \frac{1}{N} \sum_{n=2}^{N} \frac{\left|(B_n - B_{n-1}) - (T_{I_{n}} - T_{I_{n}-1}) \right|}{B_n - B_{n-1}}. \label{eq:tempo_matching_accuracy} 
\end{equation}

\subsection*{Oscillatory Tatum Dynamics Support Robust Tatum Detection in Unseen Rhythms}

The Tatum Layer is designed to provide a fine-grained temporal scaffold for the rest of the hierarchy. We first evaluated whether this layer could generalize to previouly unseen rhythmic patterns and tempi, and compared its performance with GRU and TFTransformer baselines trained under the same prediction framework. This comparison was used to assess whether the physics-inspired reservoir provides a useful temporal inductive bias for rhythm generalization.

The prediction target was shifted forward in time to evaluate anticipatory timing. This shift provides a behavioral operationalization of prediction, without implying that biological rhythm learning depends on an explicit temporal-offset supervisory signal. It allows us to test whether the model can express timing predictions before expected rhythmic events, consistent with predictive accounts of motor control~\cite{wolpert1998internal,bastian2006learning} and rhythm-specific evidence for internalized timing before expected beat onsets~\cite{fujioka2012internalized}.

\begin{figure*}[ht]
    \centering
    \includegraphics[width=\textwidth, trim={0cm 0 0cm 0cm}, clip]{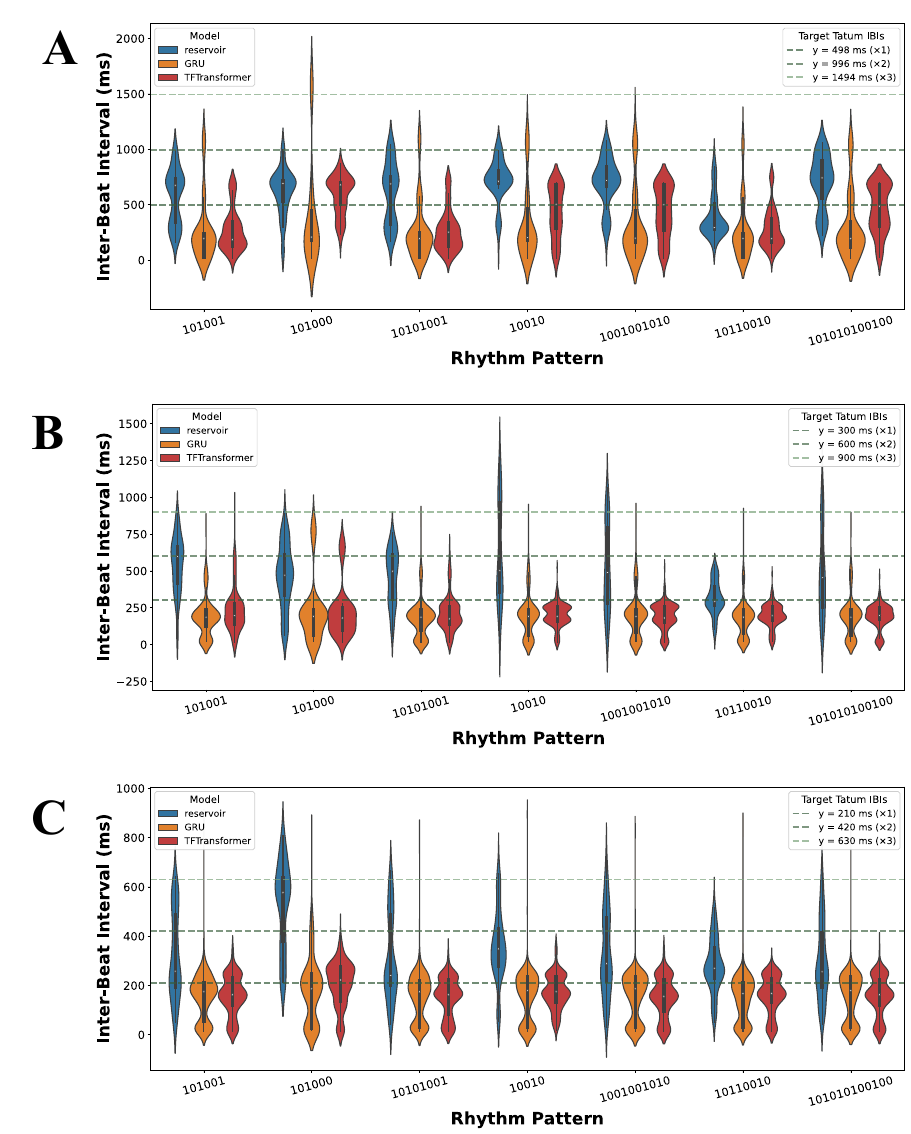}
    \caption{\textbf{Oscillatory Tatum Dynamics vs. Perceptron-Based Models in Tatum-Interval Distributions After Pretraining.} Interval distributions are shown for each model (GRU, TFTransformer, and the proposed Tatum Layer) at three different base tempi: 60~BPM, 100~BPM, and 140~BPM. The Tatum Layer exhibits interval distributions centered on target values and their integer multiples (harmonics), consistent with sensitivity to the tatum as the smallest metric unit and to its hierarchical subdivisions. GRU and Transformer models produce narrow distributions at arbitrary intermediate values, indicating learned templates that fail to lock onto fundamental tatum frequencies or their harmonics. Horizontal dashed lines indicate the target tatum interval and its integer multiples for each tempo.}
    \label{fig:4}
\end{figure*}

Fig.~\ref{fig:4} illustrates the IBI distributions produced by each model after training. In this Tatum-layer analysis, IBI denotes the interval between successive detected output pulses. While none of the models perfectly captured the exact tatum intervals (500, 300, and 214~ms), a striking contrast emerges between the resonance-based Tatum Layer and the strictly sequential GRU and TFTransformer. The latter two models exhibit limited rhythmic flexibility: they tend to lock onto a single IBI template and fail to adapt to the structural diversity of unseen rhythmic patterns. Their narrow distributions concentrate around intermediate intervals that do not correspond clearly to the target tatum or its integer multiples. In contrast, the Tatum Layer produces a broader distribution spanning tatum-related periodicities (1×, 2×, and 3× the target), reflecting greater rhythmic variability.

\begin{figure*}[ht]
    \centering
    \includegraphics[width=\textwidth, trim={0cm 0 0cm 0cm}, clip]{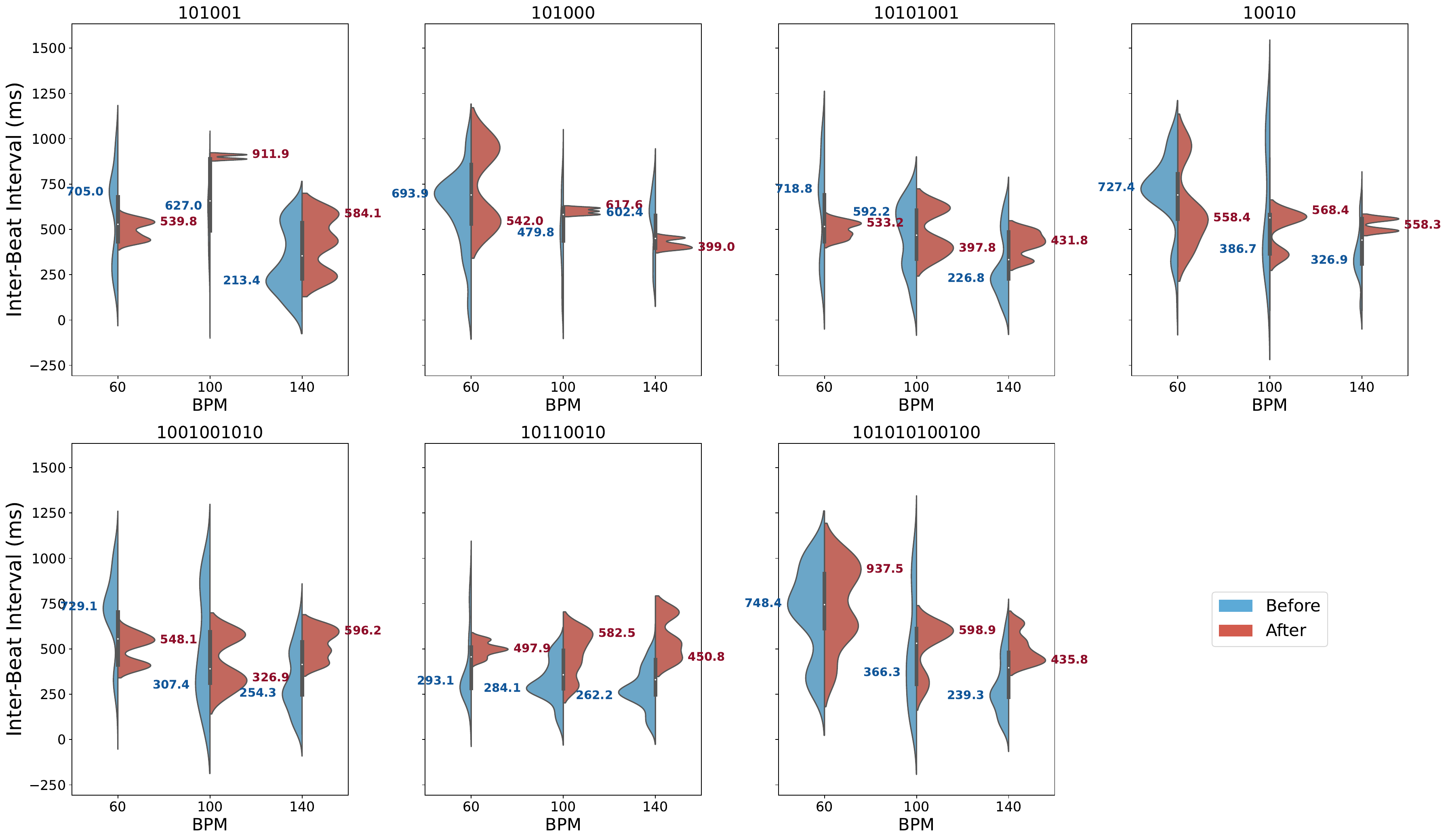}
    \caption{\textbf{Tatum Layer Exhibits Rhythmic Generalization via Oscillatory Calibration.} Violin plots show the distribution of tatum intervals produced by the Tatum Layer before and after the online calibration stage, across rhythms at three tempi. Post-calibration results show increased alignment with expected tatum intervals and their harmonics, reflecting the model's ability to generalize to novel rhythmic structures through its resonant dynamics.}
    \label{fig:5}
\end{figure*}

Fig.~\ref{fig:5} compares the Tatum Layer output before and after online calibration. After calibration, the IBI distributions show clearer alignment with the expected tatum intervals or their integer-related multiples, indicating that the fixed oscillatory reservoir can be adapted to a newly presented rhythm through readout-level calibration.

At 60~BPM, several patterns show post-calibration IBIs close to the 500~ms target or to slower integer-related multiples. For example, the pattern \texttt{101000} shows an interval near the target tatum, whereas longer or more complex patterns can show intervals closer to slower subdivisions. This suggests that the Tatum Layer can align either with the base tatum periodicity or with salient slower periodic structure, depending on the rhythmic pattern.

At 100~BPM, most patterns exhibit two prominent peaks in their IBI distributions (e.g., 397.8~ms and 616.9~ms; 326.9~ms and 580.7~ms), indicating that the model resonates at both the target tatum and higher-order multiples, likely reflecting the perceived metric structure. Even in patterns like \texttt{101001}, where the distribution centers on longer IBIs, this reflects a musically valid re-interpretation of the rhythmic grouping.

At 140~BPM, the model tends to favor slower subdivisions (double or triple tatum), consistent with human perceptual thresholds near 250~ms. For instance, in \texttt{101010100100}, peaks at 435.8~ms and 640.6~ms show how the system adapts dynamically based on tempo, allowing richer motor and cognitive entrainment.

Finally, the pattern \texttt{10010} poses a recognition challenge due to its ambiguous temporal structure alternating between double and triple tatum. Nevertheless, the Tatum Layer's oscillatory dynamics reflect this ambiguity through clear bimodal IBI peaks across all tempi.

\begin{figure*}[ht]
    \centering
    \includegraphics[width=\textwidth, trim={0cm 0 0cm 0cm}, clip]{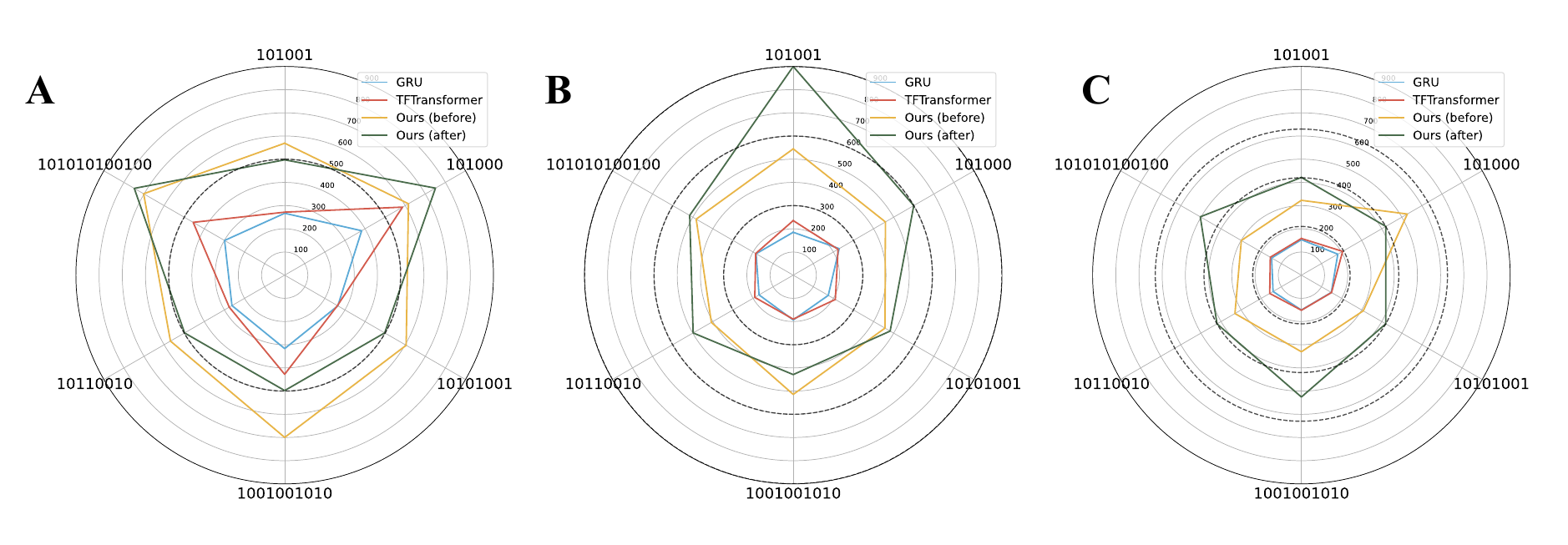}
    \caption{\textbf{Mean Tatum Intervals Reveal Resonant Responses and Rhythmic Flexibility Across Models and Tempi.} Radar plots show the mean output intervals of all rhythmic patterns at 60~BPM, 100~BPM, and 140~BPM for the GRU, TFTransformer, and the proposed Tatum Layer when each trained model is exposed to unseen rhythms. Mean interval error ratio quantifies systematic deviation: $(\text{Mean interval}_{\text{output}} - \text{interval}_{\text{target}})/\text{interval}_{\text{target}}$. Values near 0\%, 100\%, 200\% indicate entrainment to target frequency or its harmonics. Perceptron-based models (GRU and TFTransformer) produce rigid interval patterns with limited variability across input conditions, indicating poor generalization. In contrast, the oscillatory Tatum Layer demonstrates frequency-tuned flexibility and improved alignment with target tatum intervals or their harmonics under the evaluation protocol described in the Results. Black dashed hollow circles represent the target tatum values and their integer multiples.
}
    \label{fig:6}
\end{figure*}

As shown in Fig.~\ref{fig:6}, the mean IBI error ratios quantify the systematic deviation of the detected output intervals from the target tatum interval:
\[
\frac{\mathrm{Mean~IBI}_{\mathrm{output}}-\mathrm{IBI}_{\mathrm{target}}}{\mathrm{IBI}_{\mathrm{target}}}.
\]
Here, IBI is used operationally as the interval between successive detected output pulses, consistent with the figure labels. Values near 0\%, 100\%, and 200\% indicate entrainment to the target tatum interval or to its integer multiples. The mean IBI error ratios for the GRU, TFTransformer, and our Tatum Layer were 34.63\%, 25.93\%, and -50.85\%, respectively. These results show that perceptron-based models deviate from the target tatum structure and lack structural diversity in their output. In contrast, the oscillatory nature of our model allows it to generate IBI distributions that are closer to the target tatum interval or its harmonics while varying flexibly with rhythmic context.

Before online calibration, this evaluation tests responses to unseen rhythms and tempi using the pretrained readout. After online calibration, it tests rapid readout-level adaptation. While the GRU and TFTransformer failed to extrapolate meaningfully to these novel rhythms, the proposed Tatum Layer benefits from an internal frequency-tuned dynamic structure that enhances its generalization capacity.

In summary, the resonance-based dynamics of the Tatum Layer enable it to flexibly entrain to varying rhythmic structures and produce temporally coherent tatum estimates, even in unseen contexts. In contrast, GRU and TFTransformer models operate as perceptron-based sequence predictors that rely on static internal transitions and lack any intrinsic temporal variability. As a result, they tend to produce rigid IBI outputs that fail to adapt across tempi or rhythmic complexity. These limitations suggest that standard recurrent and attention-based baselines are less well suited to latent tatum inference under the present generalization setting. In this work, we therefore center our architecture around oscillatory principles, which are well suited to generalization in temporally variable settings.

Importantly, the Tatum Layer is not intended to solve the same problem as the Tactus Layer or to be judged by the same criterion of apparent output sharpness. In our model, the Tatum Layer provides a rhythm-general estimate of the smallest temporal subdivision and therefore serves as a shared temporal scaffold, whereas the Tactus Layer performs a higher-level grouping operation on top of this scaffold and is allowed rhythm-specific readout adaptation. Because the Tactus Layer operates at a slower and more stable metrical level, its outputs can appear more precise. Nevertheless, the tatum representation remains necessary because both tactus extraction and downstream movement-timing prediction must be anchored to a common fine-grained temporal reference frame.

\subsection*{Tactus-Layer Timing Relative to an Algorithmically Derived Grouping Target}
\label{subsection:tactus}

We next examined whether the Tactus Layer could organize the tatum-level scaffold, together with the surface rhythm, into slower metric groupings. The analysis focuses on whether the detected tactus intervals correspond to plausible tatum multiples and whether the resulting tactus pulses align consistently with the repeated rhythmic pattern. These results provide the intermediate metric context for interpreting the motor timing deviations analysed in the following section.

\begin{figure*}[ht]
    \centering
    \includegraphics[width=\textwidth, trim={0cm 0 0cm 0cm}, clip]{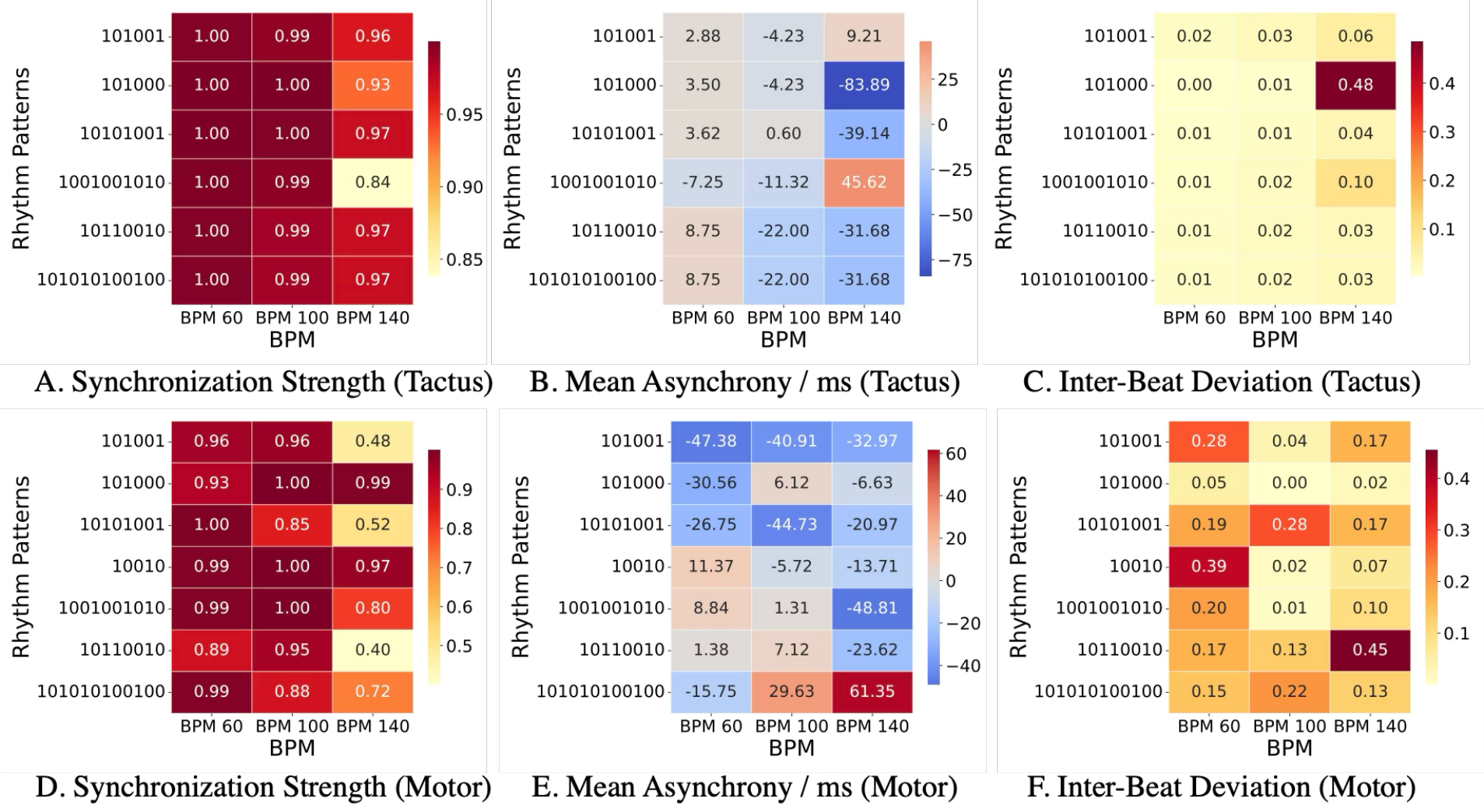}
    \caption{\textbf{Comparison of Tactus and Motor Layer Outputs Across Patterns 
and Frequencies.} 
\textit{Organization:} Rows represent seven rhythm patterns (101001, 101000, 
10101001, 10010, 1001001010, 10110010, 101010100100); columns represent three 
tempi (60, 100, 140 BPM).
\textit{Tactus Layer (Panels A, B, C):} Synchronization Strength (A, range: 0–1), 
Mean Asynchrony (B, unit: ms), and IBD (C, range: 0–1).
\textit{Motor Layer (Panels D, E, F):} Corresponding measurements—Synchronization 
Strength (D, range: 0–1), Mean Asynchrony (E, unit: ms), and IBD 
(F, range: 0–1).
Heat map colors: warmer tones indicate higher synchronization strength or lower 
timing error; cooler tones indicate lower synchronization or higher error. The 
Tactus Layer demonstrates robust phase-locking across patterns (high D values), 
while the Motor Layer shows tempo-dependent variability, reflecting the increasing 
difficulty of predicting surface-level event onsets when tactus-level expectations conflict 
with notated rhythm.
}
    \label{fig:7}
\end{figure*}

\begin{figure*}[ht]
    \centering
    \includegraphics[width=\textwidth, trim={0cm 0 0cm 0cm}, clip]{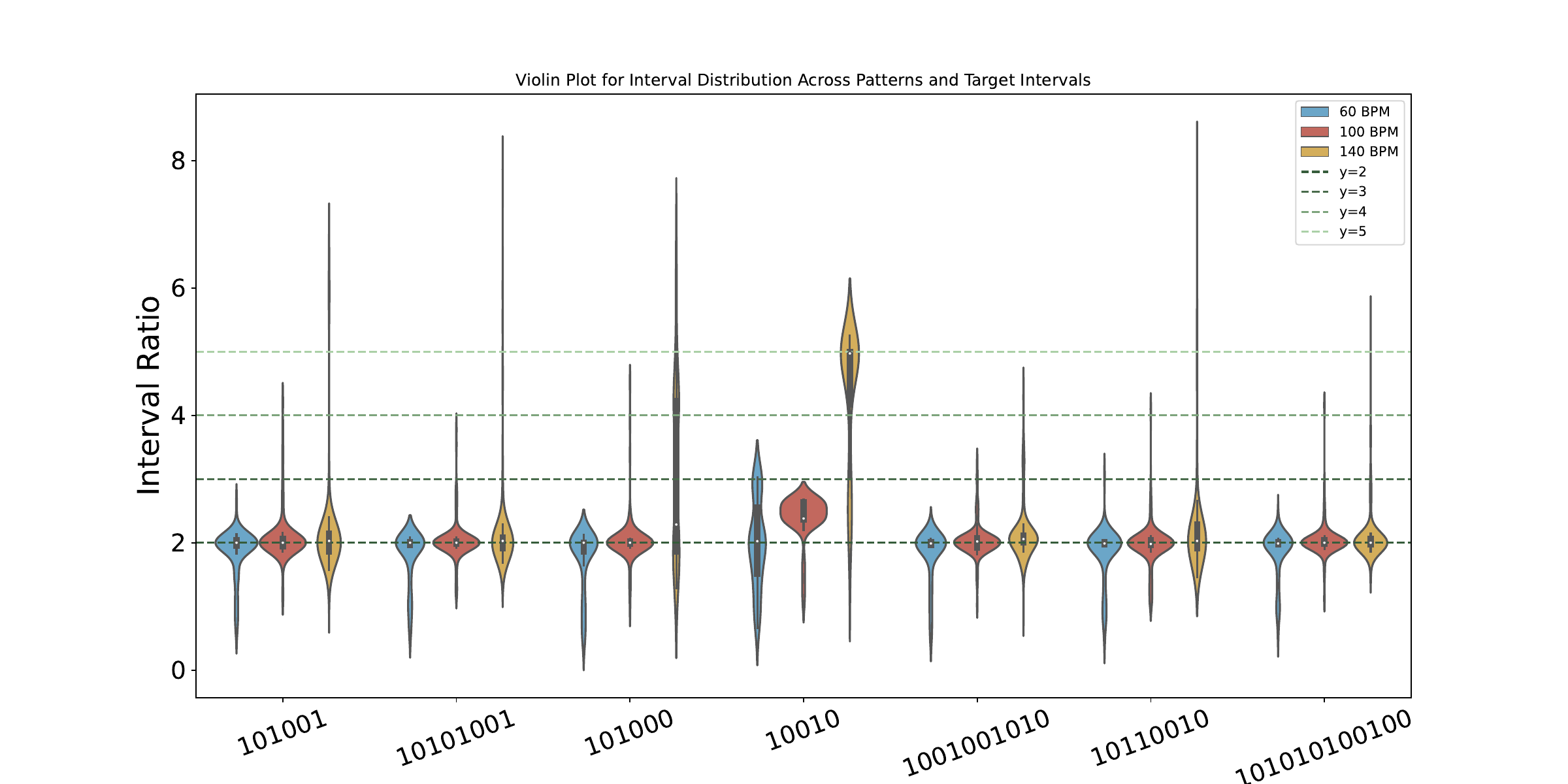}
    \caption{\textbf{Tactus Interval Distribution Aligned with the Target Smallest Time Unit.} The inter-beat intervals generated by the Tactus Layer for all rhythmic patterns, evaluated at three distinct frequencies, are normalized by their respective target tatum intervals. The y-axis represents the resulting distribution, illustrating that the Tactus Layer effectively captures the slower meter in hierarchical rhythm perception.}
    \label{fig:8}
\end{figure*}

As an overview, the layer-wise tactus and motor performance metrics are first summarized in Fig.~\ref{fig:7}, while the tactus interval distributions and circular onset visualizations are then presented in Figs.~\ref{fig:8} and \ref{fig:9}, respectively.

Tactus output peaks are identified, and the corresponding IBI values are calculated operationally as intervals between successive tactus output peaks. The IBI distributions across different patterns and beats per minute (BPM) are visualized in Fig.~\ref{fig:8}. The x-axis represents the rhythmic patterns, while the y-axis indicates the multiples of the target smallest time unit corresponding to each IBI. Horizontal reference lines at double, triple, quadruple, and quintuple multiples are included to illustrate how the tactus interval distributions vary with frequency changes.

As shown in Fig.~\ref{fig:8}, the Tactus Layer expresses binary groupings in most cases. However, the upper tails of the distributions tend to elongate as frequency increases, indicating a bias toward slower tactus levels. For example, in the pattern \texttt{101000}, at 140 BPM, the model often captures a quadruple-multiple grouping. In a more ambiguous case, \texttt{10010}, the changes in the tactus interval distribution are more pronounced. At the low frequency of 60 BPM, the distribution exhibits two distinct peaks at double and triple the target smallest time unit. As the frequency increases to 100 BPM, the tactus tends to divide the pattern length into intervals approximately 2.5 times the target unit. At 140 BPM, the model captures the entire pattern using quintuple multiples, reflecting a significant adaptation to the faster rhythm.

To elucidate the tactus and motor timing patterns, we use circular visualizations in Fig.~\ref{fig:9}. In these plots, one full rotation represents one complete repetition of the rhythmic pattern, and the uniformly spaced reference lines indicate tatum-aligned positions. Dark red dashed lines mark surface event positions, whereas light red dotted lines mark silent tatum positions. Blue dots indicate detected Tactus Layer peaks and red dots indicate detected Motor Layer output onsets. The angular position of each dot represents its phase within the rhythmic cycle, while the radial position indicates the most recent output interval, labelled as IBI in the figure. This visualization allows tactus-level grouping and motor-level timing to be compared within the same rhythmic reference frame.

Fig.~\ref{fig:9}(a, b, c) illustrate three common rhythmic patterns: \texttt{101001} and \texttt{101000}, which are three-tatum grouping patterns, and \texttt{10101001}, a two-tatum grouping pattern. Corresponding measurements of movement-timing prediction performance for these patterns are shown in the first row of Fig.~\ref{fig:7}. The Tactus and Motor Layers are evaluated against different ground truths (regular pulse vs. complex pattern), making direct metric comparison inappropriate. The Motor Layer's lower synchronization values reflect the inherent difficulty of predicting irregularly-spaced surface events. The tactus pulses are well-aligned with the target patterns, as evidenced by the synchronization strength, which exceeds 0.9 for all patterns across different frequencies. Relative phase errors are minimal, although a slight decline in synchronization strength is observed at 140 BPM.

From Fig.~\ref{fig:9}(a, b, c), the Tactus Layer output dots cluster clearly around the two-digit pattern \texttt{10}, though clusters for \texttt{101000} in Fig.~\ref{fig:9}(c) appear noisier at 140 BPM compared to other frequencies. The mean asynchrony and IBD corroborate this observation in Fig.~\ref{fig:9}(c), indicating larger distance errors in the movement-timing predictions at 140 BPM. For the overall match rate, the pattern \texttt{101000} shows lower accuracy due to occasional misalignment, with the Tactus Layer detecting a two-tatum grouping instead. This misalignment is consistent with the distribution shown in Fig.~\ref{fig:8}. Notably, larger absolute mean asynchronies are negative, indicating that the Motor Layer prediction anticipates the target event onset by occurring earlier than the target timing. These deviations are not necessarily detrimental, as they remain systematic and aligned with the rhythmic pattern.

Fig.~\ref{fig:9}(e, f, g) explore more complex patterns: \texttt{10010010}, \texttt{10110010}, and \texttt{101010100100}. Measurements of movement-timing prediction performance for these patterns are shown in the second row of Fig.~\ref{fig:7}. While the Tactus Layer output peaks are similarly well-aligned with the target patterns, synchronization strengths tend to be slightly lower than those observed for shorter grouping patterns, reflecting the increased complexity. At 140 BPM, the model exhibits greater variability, with slightly larger absolute mean asynchronies. Their predominantly negative values indicate early movement-timing predictions. The IBD values remain close to zero, confirming that the Tactus Layer reliably identifies repeated two-digit patterns such as \texttt{10}, even in more complex rhythmic scenarios. This consistency highlights the robustness of the model in synchronizing with rhythmic patterns across varying levels of difficulty.

For the hard pattern \texttt{10010}, the tactus exhibits greater ambiguity, with tactus intervals that are not consistently equal. As shown in Fig.~\ref{fig:9}(d), at slower tempos, the tactus pulses tend to cluster around double and triple tatums. Additionally, other clusters emerge at positions suggesting ternary subdivision (three roughly equal intervals)—a documented grouping preference in human rhythm perception \cite{povel1985perception, grube2010dissociation}. The other frequencies results shown in the figure is well aligned with the numerical results.

\subsection*{Movement-Timing Prediction in the Motor Layer}

The Motor Layer generates a movement-timing prediction by synthesizing inputs from both the Tatum and Tactus Layers. It integrates the tatum-level scaffold and tactus-level grouping into a predicted movement-timing signal; using the circular visualizations introduced above, we examined how this prediction aligned with surface event positions and how tactus-related expectations influenced its timing deviations.

\begin{figure*}[ht]
    \centering
    \includegraphics[width=\textwidth, trim={0cm 0 0cm 0cm}, clip]{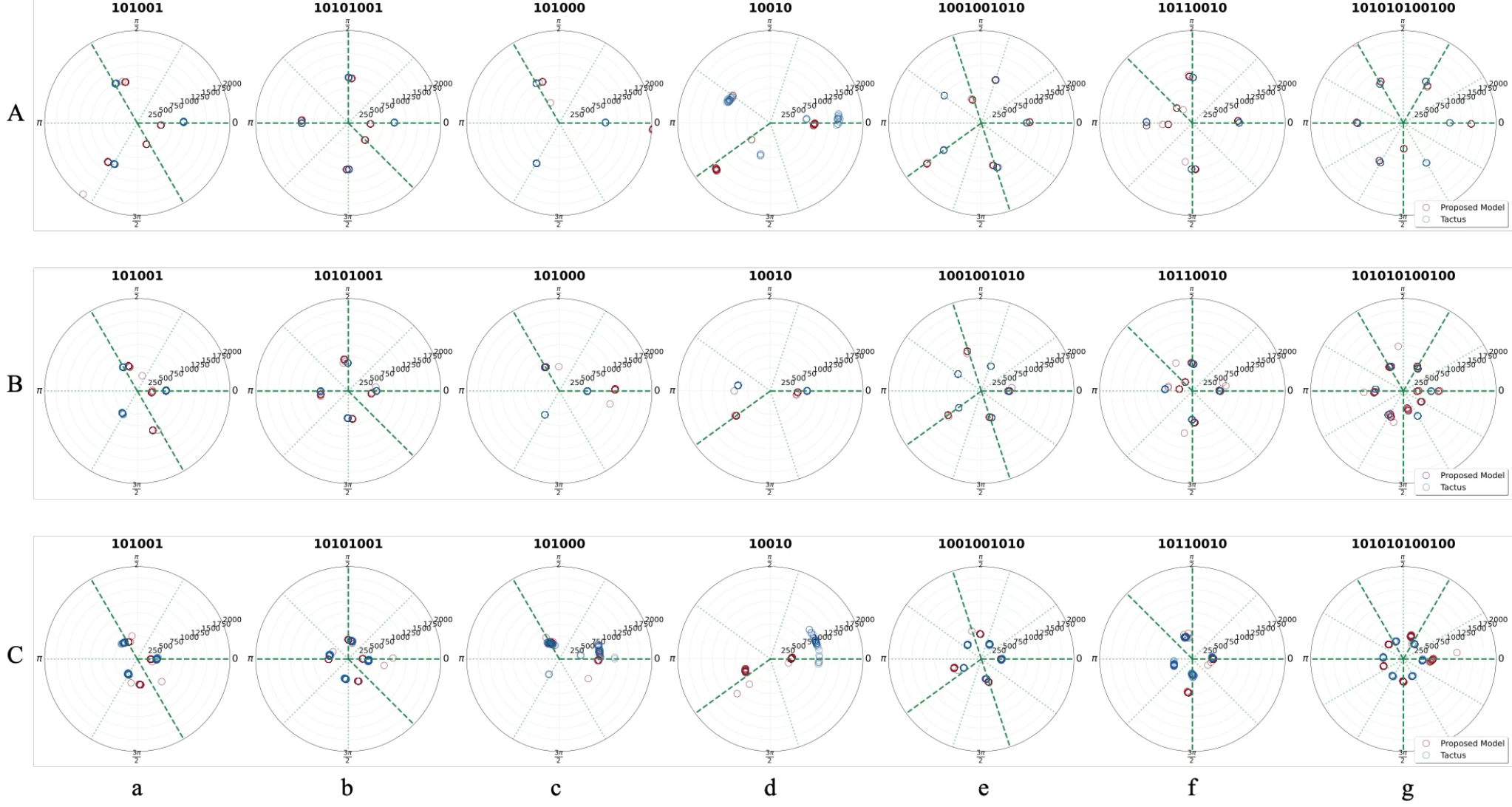}
    \caption{\textbf{Visualization of Rhythmic Patterns Highlighting the Alignment 
of Tactus Layer and Motor Layer Output Onsets Across the Entire Sequence.}
\textit{Layout:} Rows (A, B, C) = tempi (60, 100, 140 BPM); columns (a–g) = 
rhythm patterns (101001, 10101001, 101000, 10010, 1001001010, 10110010, 
101010100100).
\textit{Circular encoding:} Angular position = phase shift within the metric cycle; 
radial distance = inter-beat interval. Blue circles = Tactus Layer peaks; red 
circles = Motor Layer peaks. Dark red dashed lines = event-onset positions (`1'); light red 
dotted lines = silent positions (`0').}
    \label{fig:9}
\end{figure*}

The onset visualizations further illustrate the alignment between the Motor Layer predictions and the target rhythmic patterns. Fig.~\ref{fig:9} presents seven representative patterns, while the corresponding measurements of movement-timing prediction performance for these patterns are summarized in Fig.~\ref{fig:7}. The results demonstrate that the Motor Layer maintains consistent phase relationships with these patterns (R = 0.85-0.95), though with some systematic timing deviations (mean asynchrony $\approx$ -40 ms) and occasional extra peaks induced by tactus-motor conflicts, particularly at faster tempi.

The mean asynchronies reported in Fig.~\ref{fig:7} are approximately -40 ms, with predominantly negative values, indicating that the model slightly anticipates the target event onset by producing outputs marginally ahead of the surface event onset. This behaviour aligns with observations in embodied systems, where such anticipation is a common phenomenon.

Given the variability in the Motor Layer predictions, identifying overarching trends is challenging. To address this, we analyze each rhythmic pattern individually, focusing on frequency-dependent variations.

For the pattern \texttt{101001}, shown in Fig.~\ref{fig:9}(a), synchronization strength at 140BPM is the lowest among the three tested frequencies when considering only predicted movement onsets that align correctly with the pattern. As illustrated in Fig.~\ref{fig:9}(A.a), the Motor Layer prediction onsets synchronize well with the first two event onsets but lead the timing for the third event. Additionally, Fig.~\ref{fig:9}(C.a) shows an extra peak in the Motor Layer prediction, caused by the tactus pulse not aligned with the event-onset positions in the pattern. While pattern 101001 is metrically regular, at 140 BPM the tatum (214 ms) approaches the lower limit of robust beat perception \cite{repp2005sensorimotor}, and the extra peak may reflect tempo-dependent perceptual ambiguity. This misalignment drifts the third event off its correct timing. The mean asynchrony remains consistently negative across frequencies without transitioning to another digit. Notably, the IBD value at 60~BPM is the highest, attributed to the extra peak generated by the Motor--Tactus Layer interaction.

For the pattern \texttt{101000}, shown in Fig.~\ref{fig:9}(c), synchronization strength exceeds 0.9 for all tested frequencies, considering only properly matched Motor Layer prediction pulses. The mean asynchrony values are consistently small and negative, with the largest being $-30.56$. All IBD values are below 0.05, indicating no extra peaks generated by the Motor Layer prediction. This observation aligns with the prediction pattern depicted in Fig.~\ref{fig:9}(c).

For the pattern \texttt{10101001}, shown in Fig.~\ref{fig:9}(b), the behaviour is similar to pattern \texttt{101001}. Synchronization strength at 140BPM is the lowest among the frequencies tested, while other frequencies exhibit stronger synchronization. From Fig.~\ref{fig:9}(A.b), the Motor Layer prediction onsets match the first two event onsets but lead the timing for the third event. As shown in Fig.~\ref{fig:9}(C.b), the Motor Layer prediction contains an additional peak, driven by the tactus pulse that does not coincide with the event-onset positions in the pattern. Consequently, the third event drifts away from its proper timing. The mean asynchrony remains negative across all frequencies, while the IBD value at 60~BPM is the highest, driven by the extra prediction peak induced by the Motor--Tactus Layer interaction.

For the pattern \texttt{1001001010} shown in Fig.~\ref{fig:9}(e), the synchronization strength remains consistently above 0.8 across all tested frequencies. However, at 140 BPM, the synchronization strength is the lowest, while other frequencies exhibit stronger synchronization. The mean asynchrony also shows a larger error at 140 BPM, supporting the conclusion derived from the synchronization strength measurement. For the IBD values, an example at 60 BPM shows a value of 0.1999, indicating the generation of extra peaks by the model. Specifically, as illustrated in Fig.~\ref{fig:9}(A.e), the Motor Layer prediction contains an extra peak at the third digit of the pattern, driven by the tactus.

For the pattern \texttt{10110010} shown in Fig.~\ref{fig:9}(f), the synchronization strength remains consistently above 0.88 at lower frequencies. The tactus is reliably captured; however, the third ‘1’ in the pattern does not align with the tactus, leading to an ambiguous movement-timing prediction at this position. The nature of the prediction varies across different tempi. At 60 and 100 BPM, the synchronization strength approaches 1, and the mean asynchrony is smaller compared to higher tempi. However, the IBD values exceed 0.1. As illustrated in Fig.~\ref{fig:9}(A.f) and Fig.~\ref{fig:9}(B.f), Motor Layer prediction pulses are omitted at the fourth digit in the pattern, a phenomenon that aligns with the observed tactus. At faster tempi, such as 140 BPM, the prediction pulses at the ambiguous position exhibit noticeable drift. As shown in Fig.~\ref{fig:9}(C.f), the red dots deviate from the second dark red line and cluster around the tactus pulse. This drift is attributed to the difficulty in distinguishing the two consecutive ‘1’s at higher speeds, causing the model to omit one. Consequently, correct movement-timing prediction pulses are infrequently generated at the third ‘1’ in the pattern, resulting in increased mean asynchrony and larger IBD values.

For the pattern \texttt{101010100100} shown in Fig.~\ref{fig:9}(g), the synchronization strength decreases as the BPM increases. This trend likely reflects two compounding factors: the relatively long cycle 
length of this pattern (12 tatums), which places greater demands on the model's temporal memory for maintaining stable hierarchical predictions, and the shorter tatum intervals at faster tempi, which increase metric ambiguity as the tatum approaches the lower limit of beat perception (200--250 ms) \cite{repp2005sensorimotor}. At 60 BPM, the synchronization strength is nearly 1, and the mean asynchrony is -15.75 ms, indicating that the Motor Layer prediction occurs slightly before the target event onset. However, the IBD value is 0.1479 due to an extra peak at the ninth digit, as shown in Fig.~\ref{fig:9}(A.g). These results suggest that the Tatum Layer captures 12 tatum peaks for this pattern, simplifying the prediction task for the higher cognitive layers. At 100 and 140 BPM, the variance increases, with mean delays of 29.6308 ms and 61.3548 ms, respectively. The clusters in Fig.~\ref{fig:9}(B.g) and Fig.~\ref{fig:9}(C.g) are noisier than those at 60 BPM. At 100 BPM, the IBD value increases to 0.2154 due to ambiguous tatum representation, as the Tatum Layer fails to capture all 12 tatum peaks and instead combines some smaller time units. At 140 BPM, while the model matches all target event onsets, it exhibits a mean delay of 61.3548 ms, reflecting a degradation in timing precision.

For the challenging pattern \texttt{10010}, the results are presented in Fig.~\ref{fig:7} and Fig.~\ref{fig:9}(d). The Motor Layer prediction pulses closely aligned with the pattern event onsets. However, additional peaks induced by the Tactus Layer are evident, particularly between the third and fourth digits in Fig.~\ref{fig:9}(A.d) and Fig.~\ref{fig:9}(B.d), where a cluster of tactus pulses emerges.

These tactus-induced timing deviations observed in our model reflect a fundamental challenge in human rhythmic perception and motor synchronization. When the tactus layer captures a slower metric level that does not perfectly align with the surface rhythm's event onsets, it creates competing temporal expectations that influence the model's movement-timing predictions. This phenomenon resembles documented tactus-metric conflicts in human rhythm perception \cite{keller2005staying, fitch2007perception, mathias2020rhythm}, though systematic validation requires collecting human behavioral data on our specific patterns to test whether humans also generate spurious taps at tactus positions when surface rhythm and perceived pulse misalign. In human studies, such conflicts can produce anticipatory motor responses that drift toward the internal pulse. The negative mean asynchronies (anticipatory predictions) combined with occasional extra predicted movement onsets in our model parallel the timing errors frequently observed in human tapping studies, particularly when participants must synchronize with rhythms that challenge their internal metric representations.

\subsection*{Qualitative Motor Layer Beta-Band Activity}

We analyzed beta-band (13–30 Hz) activity in the Motor Layer as an exploratory qualitative comparison with rhythm-sensitive beta activity reported in human EEG/MEG studies~\cite{merchant2015finding,fujioka2012internalized,fujioka2015beta}. The model was not fitted to neural recordings, so this analysis is used only to characterize whether Motor Layer activity contains rhythm- and tempo-dependent beta-band structure.

The beta-band activity analysed here reflects Motor Layer dynamics under Higher Cognition-guided suppression. The Higher Cognition Layer predicts event/silence labels at tatum-aligned positions, and its silent-position predictions provide the top-down control signal that suppresses Motor Layer output at learned silent positions. Thus, beta-band activity is measured in the Motor Layer during this suppression condition; it is not treated as a direct output of the Higher Cognition Layer.

\begin{figure*}[ht]
    \centering
    \includegraphics[width=\textwidth, trim={0cm 0 0cm 0cm}, clip]{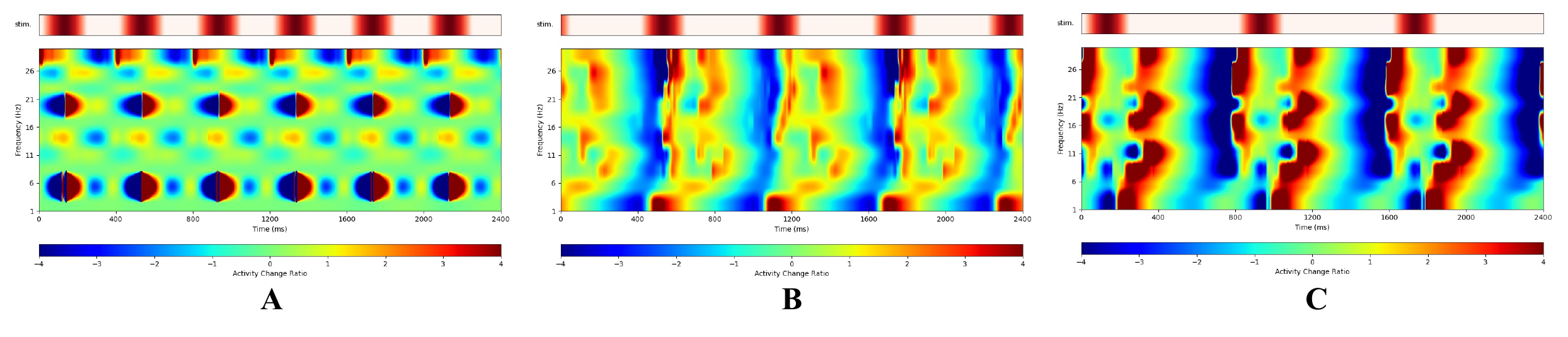}
    \caption{\textbf{Neural Activity in Motor Layer Neurons.} Hidden states of all $p$ neurons were recorded while processing rhythmic sequences with increasing IBIs (400, 600, and 800\,ms), where IBI denotes the interval between successive surface-event onsets in this figure. Neural envelopes were extracted via Hilbert transform (0--30\,Hz, 3\,Hz bins), following the analytic method used in human MEG studies~\cite{fujioka2012internalized}. Beta-band (13--30\,Hz) activity shows IBI-dependent local modulation around event-related and imagined tatum-level pulse positions. In the longest-IBI condition, additional local beta increases appear around imagined intermediate tatum-level pulse positions, consistent with an internal counting-like process across long gaps. Surface-event markers are overlaid above each spectrogram.}
    \label{fig:10}
\end{figure*}

Across the tested rhythms and tempi, Motor Layer beta-band activity varies around event-related and tatum-aligned positions. In Fig.~\ref{fig:10}, the beta-band signal shows rhythm- and tempo-dependent local modulation near these positions. The relevant observation is that the timing of these local beta-band fluctuations changes with rhythm pattern and tempo.

\begin{figure*}[ht]
    \centering
    \includegraphics[width=\textwidth, trim={0cm 0 0cm 0cm}, clip]{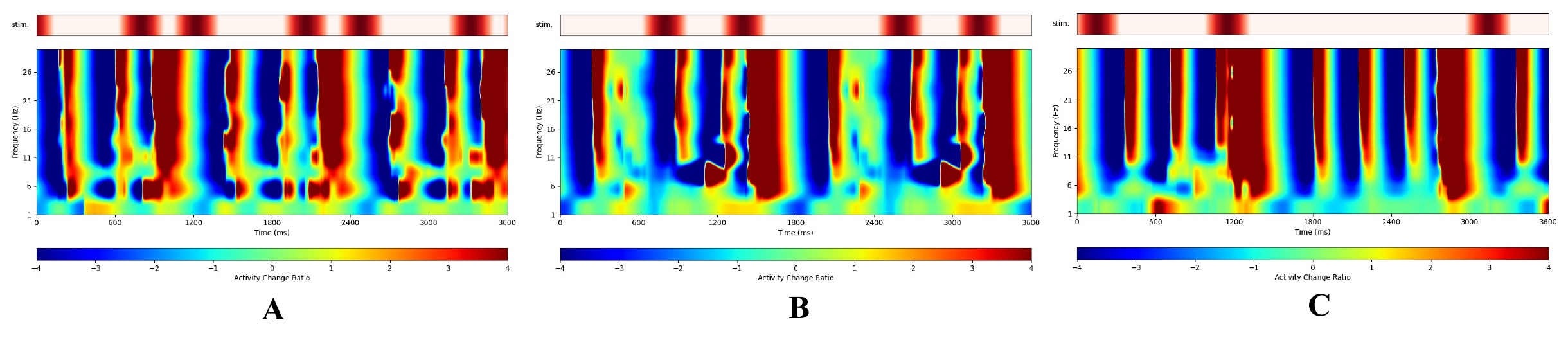}
    \caption{\textbf{Neural Activity in Motor Layer Neurons for Pattern \texttt{101000} Across Tempi.} The model processed the syncopated pattern \texttt{101000} at 140, 100, and 60 BPM. Neural envelopes were extracted via Hilbert transform (0--30 Hz, 3 Hz bins), following EEG analysis methods in~\cite{fujioka2012internalized,fujioka2015beta}. The pattern can be parsed as a repeated \texttt{10} grouping followed by two silent tatum positions (\texttt{10 10 00}). Beta-band (13--30 Hz) activity shows tempo-dependent local modulation near imagined tatum-level pulse positions, while Motor Layer output is suppressed at the learned silent positions. Surface-event markers are shown above each spectrum.}
    \label{fig:11}
\end{figure*}

Fig.~\ref{fig:11} illustrates the same point for the pattern \texttt{101000}. In this case, Motor Layer output is suppressed at learned silent positions while beta-band activity continues to vary around the corresponding tatum-aligned positions. The timing and direction of these beta-band changes differ across tempi, so we describe the result as rhythm- and tempo-dependent beta-band modulation under Higher Cognition-guided suppression.

Overall, the beta analysis is best interpreted as an exploratory observation. Motor Layer beta-band activity changes with rhythm pattern and tempo, including when Motor Layer output is suppressed by the Higher Cognition Layer, providing qualitative evidence for rhythm-sensitive internal dynamics in the model.
\section*{Discussion}

Understanding how humans perceive and anticipate complex rhythmic patterns remains a fundamental challenge in cognitive neuroscience and artificial intelligence. While previous research has largely focused on metronome-based tasks, the mechanisms underlying human perception of hierarchical musical rhythms—particularly the systematic errors and timing deviations that characterize natural rhythmic behavior—remain poorly understood. To address this gap, we developed a biologically inspired computational model that tests whether physics-based oscillatory dynamics can reproduce selected components of hierarchical rhythm processing.

Our main contribution is to show that a shared oscillatory reservoir, combined with layer-specific readout calibration to each new rhythm, can support several components of hierarchical rhythm processing. The Tatum Layer provides the fine temporal scaffold, the Tactus Layer introduces an intermediate metric grouping, the Higher Cognition Layer supplies event–silence context, and the Motor Layer expresses these signals as movement-timing predictions. This division is important because the model output cannot be explained by a Tatum-to-Higher-Cognition pathway alone: tactus-like grouping creates intermediate timing expectations that can interact with surface event positions and shape deviations in predicted movement timing.

A central insight of this work is that the wave-based reservoir dynamics provide the mechanistic foundation for the model behaviors described above. The spatial gradient in wave speed $c[i,j]$ across the FDTD grid endows each layer with a structured family of resonant modes spanning a continuous range of frequencies: the slow end of the grid naturally supports low-frequency oscillations corresponding to tactus and measure-level periodicities, while the fast end captures tatum-level dynamics. The fixed reservoir supplies this frequency hierarchy, and the readouts select and combine its modes through the calibration procedures described in the Methods.

The wave dynamics also provide a principled mechanism for switching between entrainment and suppression of movement-timing predictions without weight updates. As established in the Theoretical Framework, changing the sign of the damping parameter $k^p$ shifts the Motor Layer from stable oscillation to an unstable output-suppression regime. A scalar control signal from the Higher Cognition Layer can therefore suppress a prediction at an anticipated-but-silent tatum-level pulse. This transient control affects all dynamical modes and produces a computational reduction of the Motor Layer prediction. This mechanism should be interpreted as a model-specific implementation of top-down control. We relate this mechanism only broadly to human evidence that sensorimotor beta activity is modulated during rhythmic timing tasks~\cite{fujioka2015beta}.

The generalization results support this mode-selection account. In Figs.~\ref{fig:4} and \ref{fig:6}, the GRU and TFTransformer converge to intermediate IBI values that do not clearly correspond to the target tatum or its integer multiples, whereas the Tatum Layer produces intervals closer to the target tatum or its harmonics. This contrast suggests that the structured resonant basis helps the model adapt across rhythms and tempi through brief readout calibration.

The calibrated tactus-like grouping also shapes movement-timing predictions: when it conflicts with surface-event positions, the Motor Layer produces systematic timing deviations resembling those reported for tactus--surface conflicts in human rhythm processing~\cite{keller2005staying,fitch2007perception,mathias2020rhythm}. Because the Tactus Layer is calibrated using an algorithmically derived grouping target, these results demonstrate how an imposed intermediate representation can influence downstream timing predictions; they do not demonstrate autonomous tactus or meter induction.

Motivated by human EEG/MEG studies showing that beta activity is modulated by rhythmic timing, metrical structure, and imagery~\cite{fujioka2012internalized,fujioka2015beta,iversen2009top}, we analyzed beta-band activity in the Motor Layer as an exploratory qualitative comparison. The model shows rhythm- and tempo-dependent local beta-band modulation around event-related and tatum-aligned positions, including during Higher Cognition-guided suppression of Motor Layer output. These observations suggest that the Motor Layer contains rhythm-sensitive internal dynamics. Beta-band findings are strongly task- and circuit-dependent; for example, subthalamic beta activity during rhythmic finger tapping in Parkinson’s disease reflects a different motor context from the EEG/MEG timing and imagery studies considered here~\cite{joundi2013persistent}. We therefore interpret the model beta-band activity as qualitative biological plausibility evidence.

The online adaptation of $W_{\text{out}}$ similarly maps onto rapid error-based cerebellar learning, where climbing fiber signals drive pulse-by-pulse Purkinje cell weight modification during novel rhythm acquisition~\cite{bastian2006learning}. The reservoir provides a fixed temporal basis analogous to stable mossy fiber inputs, while $W_{\text{out}}$ adaptation optimizes the linear readout over this basis, functionally similar to Purkinje cell synaptic modification.

In summary, the architecture provides a computational account of how a fixed oscillatory substrate and rhythm-specific readout calibration can jointly support hierarchical timing predictions and their suppression at predicted silent positions.

The present study has several limitations. First, evaluation is restricted to a selected set of repeating binary rhythms and three tempi, so generalization to a broader range of musical structures remains to be established. Second, the tactus-like grouping is derived algorithmically from each symbolic rhythm and imposed through readout calibration; the results therefore do not demonstrate autonomous tactus or meter induction. Third, calibration relies on layer-specific supervised targets, including a fixed 200~ms anticipatory horizon for the Tatum and Motor layers, and should not be interpreted as a biological account of how these targets are learned. Finally, the model has not been fitted to or validated against behavioral or neurophysiological data for the tested rhythms. The Motor Layer output is a movement-timing prediction, and the beta-band analysis is an exploratory qualitative comparison.

Although simplified, the present framework provides a computational platform for testing mechanistic hypotheses about hierarchical rhythm perception and may serve as a basis for future comparisons with behavioral and neurophysiological data.

\section*{Methods}
\begin{figure*}[ht]
    \centering
    \includegraphics[width=\textwidth, trim={0cm 0 0cm 0cm}, clip]{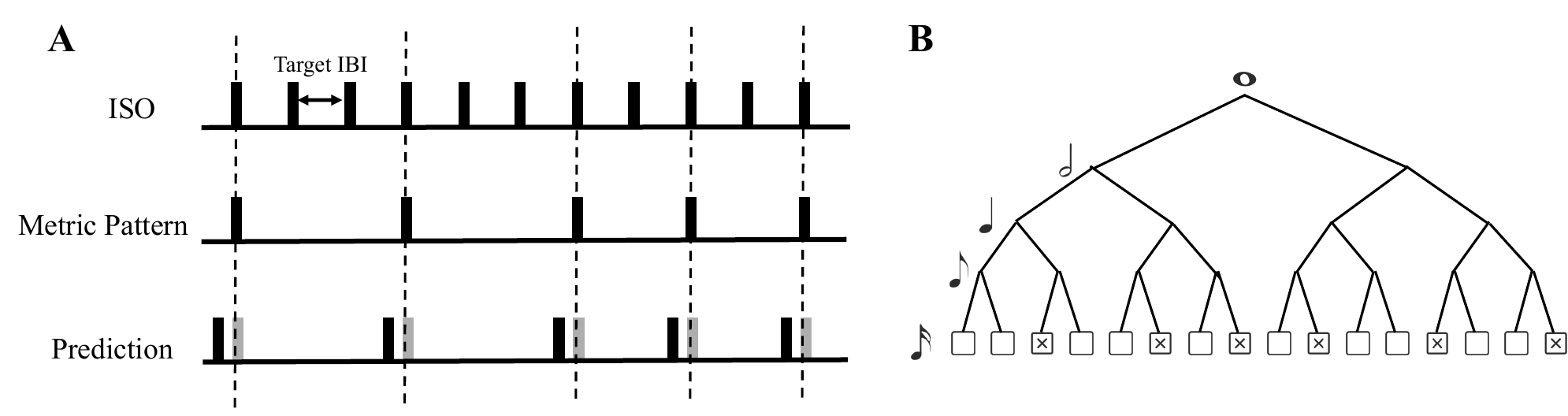}
    \caption{\textbf{Schematic representation of rhythm structure.} (A) Definitions used in the rhythm illustration. All complex patterns are derived from an isochronous (ISO) rhythm, characterized by equal pulse intervals, with the target tatum interval serving as the smallest time unit in the complex pattern. The complex pattern is a metric pattern, where each event onset aligns with a tatum position in the ISO scaffold. The 200~ms anticipatory offset is introduced by Tatum Layer pretraining and used as the shared reference frame for layer-specific online calibration. (B) Hierarchical musical perception structure, as described by Vuust et al.~\cite{vuust2014rhythmic}, is illustrated. Each metric level is recursively subdivided into equally spaced sub-units at the next lower level, defining the metric salience of positions within the rhythmic framework. The tactus lies at the midpoint of this structure.}
    \label{fig:1}
\end{figure*}

In this paper, we introduce the concept of rhythm within the context of isochronic and metrically complex patterns (see Fig.~\ref{fig:1}(A)). Isochronic rhythms are characterized by a fixed basic inter-beat interval, whereas metric rhythms exhibit a hierarchical structure that captures the subtleties of more complex rhythmic patterns, as illustrated in Fig.~\ref{fig:1}(B). We selected these rhythmic patterns at the measure level, which repeat periodically throughout the trial. The movement-timing prediction task evaluates whether the calibrated hierarchy can produce a Motor Layer prediction approximately 200~ms before the corresponding event timing; the layer-wise construction of this anticipatory reference frame is described in the Online calibration protocol.

We implemented a four-layer neural network architecture to perform the complex pattern predictive coding task, as shown in Fig.~\ref{fig:2}. The network comprises the Tatum Layer, Tactus Layer, Higher Cognition Layer, and Motor Layer. Each layer is responsible for distinct aspects of rhythm processing and prediction. The Tatum Layer detects the smallest rhythmic unit, the tatum interval, while the Tactus Layer identifies the slower underlying rhythm. The Higher Cognition Layer serves as working memory, and the Motor Layer generates motor responses based on predictions informed by working memory.

\subsection*{Layer-specific architecture and calibration workflow}

The connection structure between input and output weights varies across layers, allowing each layer to generate distinct outputs while remaining synchronized with the rhythmic input. This layer-wise differentiation enhances rhythm-based predictions. The full learning algorithm is provided in Algorithm~\ref{alg:learning_algorithm}.

\setlength{\intextsep}{5pt}
\setlength{\textfloatsep}{5pt}
\renewcommand{\baselinestretch}{0.95}

\begin{algorithm}[htbp]
\caption{Learn to predict a complex rhythm pattern}
\label{alg:learning_algorithm}
\footnotesize
\begin{algorithmic}[1]
\State \textbf{Init:} pattern\_sequence, trigger\_sequence, epochs $N$, sequence\_length $T$, learning rate $\gamma$
\State Load pretrained Tatum readout $W_{out}^{tatum}$ from the synthetic rhythm dataset; allow only an initial 0--6 s readout-level online calibration for the new rhythm.
\State Define layer-specific calibration windows $\mathcal{U}_{tatum}$, $\mathcal{U}_{tactus}$, and $\mathcal{U}_{motor}$; in the current implementation these correspond approximately to 0--6 s, 12--18 s, and 12--24 s, respectively.
\vspace{2pt}
\For{$n = 1$ to $N$}
  \For{$l = 1$ to $|\text{pattern\_sequence}|$}
    \State $\hat{y}^{hc}_{l} = \sigma(W_{out}^{hc} h^{hc}_{l})$
    \State $\mathcal{L}_{BCE} = -\frac{1}{N} \sum_{l=1}^{N} \big[y_l \log(\hat{y}_l) + (1 - y_l)\log(1 - \hat{y}_l)\big]$
    \State $W_{out}^{hc} \leftarrow W_{out}^{hc} - \text{ADAM}(\nabla \mathcal{L}_{BCE})$
  \EndFor
\EndFor

\vspace{2pt}
\For{$t = 1$ to $T$}
  \For{$* \in \{\text{tatum}, \text{tactus}\}$}
    \State $h_t^* \leftarrow (1{-}\alpha)h_{t-1}^* + \alpha f(W_{in}^* x_t^* + W^* h_{t-1} + \xi_t^*)$
  \EndFor

  \If{$t \geq$ start\_time}
    \State $W^{motor} \leftarrow \begin{cases}
      W^{motor\_damp} & \text{if } \hat{y}^{motor}=1 \\
      W^{motor\_ori} & \text{otherwise}
    \end{cases}$
  \EndIf

  \State $x_t^{motor} \leftarrow W_{out}^{tatum} h_t^{tatum} + W_{out}^{tactus} h_t^{tactus}$
  \State $h_t^{motor} \leftarrow (1{-}\alpha) h_{t-1}^{motor} + \alpha f(x_t^{motor} + W^{motor} h_{t-1} + \xi_t^{motor})$

  \If{$t \in \mathcal{U}_{tatum}$}
    \For{$i = 1$ to iterative\_step}
      \State $l_{MSE}^{tatum} = \frac{(\hat{y}^{tatum} - y^{tatum})^2}{\text{update\_step}}$
      \State $W_{out}^{tatum} \leftarrow W_{out}^{tatum} - \gamma \cdot \partial l_{MSE}^{tatum} / \partial W_{out}^{tatum}$
    \EndFor
  \EndIf
  \If{$t \in \mathcal{U}_{tactus}$ or $t \in \mathcal{U}_{motor}$}
    \For{$i = 1$ to iterative\_step}
      \For{$* \in \{\text{tactus}, \text{motor}\}$}
        \If{$t \in \mathcal{U}_{*}$}
          \State $l_{MSE}^{*} = \frac{(\hat{y}^{*} - y^{*})^2}{\text{update\_step}}$
          \State $W_{out}^{*} \leftarrow W_{out}^{*} - \gamma \cdot \partial l_{MSE}^{*} / \partial W_{out}^{*}$
        \EndIf
      \EndFor
    \EndFor
  \EndIf

  \For{$* \in \{\text{tatum}, \text{tactus}, \text{motor}\}$}
    \State $\hat{y}_t^* \leftarrow W_{out}^{*} h_t^*$
    \State \textbf{Output onset detection from} $\hat{y}^*_t$:
    \State \quad Identify candidates: $\mathcal{C} = \{t : \hat{y}^*_t > \tau\}$,
       where $\tau = 5$ (or $10$ for \texttt{101010100100})
    \State \quad Accept peak at $t^* \in \mathcal{C}$ if $\hat{y}^*_{t^*} =
       \max_{|t - t^*| \leq 30} \hat{y}^*_t$ and $t^* - t^*_{\text{prev}}
       \geq 30$
  \EndFor
\EndFor
\end{algorithmic}
\end{algorithm}

\subsection*{Reservoir structure}
\label{subsection:reservoir_structure}
All layers operate within a unified structural framework governed by the same core set of equations:
\begin{equation}
\begin{aligned}
    \mathbf{h}_{t} &= (1-\alpha)\mathbf{h}_{t-1}\\
    &+\alpha f(\mathbf{W}_{in}\mathbf{x}_{t}+\mathbf{W}\mathbf{h}_{t-1}+\xi_{t}), \\
    \hat{\mathbf{y}}_{t+\Delta t} &= \mathbf{W}_{out}\mathbf{h}_{t},
\end{aligned}
\label{eq:model}
\end{equation}
where $\mathbf{W}$ denotes the sparse recurrent connectivity, $\mathbf{W}_{in}$ and $\mathbf{W}_{out}$ are the input and readout weights, $\alpha = 0.03$ is the leakage rate, which is used uniformly across all layers, and $f(\cdot)$ is a nonlinear activation function ($\tanh$ in this work). $\mathbf{x}_t$ and $\mathbf{h}_t$ represent the input and hidden state at time $t$, respectively. The predictive horizon is $\Delta t = 200$~ms, corresponding to $n = \lfloor \Delta t / \delta t \rfloor = 33$ discrete simulation steps for $\delta t = 6$~ms; how this horizon is used during layer-wise calibration is specified in the Online calibration protocol. We set $\delta t = 6$ ms to provide sufficient temporal resolution for the rhythmic stimuli while keeping sequence lengths computationally manageable.

The symbolic rhythm pattern (e.g., \texttt{101000}) is translated into a continuous input signal $x_t$ through a two-step process. First, each symbol in the pattern is mapped to a single-timestep pulse, where \texttt{1} indicates an event onset and \texttt{0} a silent position. 
Second, each single-timestep pulse is replaced by a symmetric sine-shaped pulse of half-width $w = 30$ timesteps, giving a total pulse duration of
\begin{equation}
    \text{pulse\_duration} = (2w + 1)\delta t = 61 \times 6 = 366 \text{ ms}
\end{equation}
where the pulse amplitude follows $A \cdot \sin(0.5x)$, $x \in [0, 2\pi]$, with $A = 85$, producing a smooth half-period sinusoidal envelope centered on the event onset.

In this work, $\mathbf{W}$ is determined by the FDTD discretization. The Tatum and Tactus input matrices $\mathbf{W}_{\mathrm{in}}$ are randomly initialized once and then held fixed. The Motor Layer receives the summed Tatum and Tactus outputs directly, and the Higher Cognition Layer is driven by Tatum-output peak triggers; neither layer uses an input matrix. The term $\xi_t\sim\mathcal{N}(0,\sigma^2)$ denotes additive Gaussian noise sampled independently at each time step.

\begin{equation}
\begin{aligned}
    \frac{\partial p}{\partial t}+ k^p_{i,j} p + c^2_{i,j} \nabla \cdot \mathbf{o} &= 0, \\
    \frac{\partial \mathbf{o}}{\partial t} + k^o_{i,j} \cdot \mathbf{o} + \nabla p &= 0,
\end{aligned}
\label{eq:wave_eq}
\end{equation}
where $c_{i,j}$, $k^p_{i,j}$, and $k^o_{i,j}$ are spatially varying wave speed, pressure damping, and velocity damping at grid position $(i,j)$, respectively. For brevity, we write $c$, $k^p$, $k^o$ when referring to these fields collectively.

We emphasize that the 2D FDTD pressure/velocity system is not intended as a literal anatomical model of cortical tissue. Rather, it is a computational abstraction motivated by empirical evidence that beta-band oscillations (15–30 Hz) in human motor cortex propagate as spatially organized planar traveling waves~\cite{takahashi2011propagating}, a pattern widespread across cortical areas and species~\cite{muller2018cortical}, and accountable by spatially structured local connectivity analogous to the FDTD spatial coupling~\cite{kang2023beta}. The wave-based dynamics therefore constitute a mechanistic hypothesis warranting future empirical testing, not an established neural fact.

While $W$ is structured according to the FDTD discretization, central differences, and explicit time-stepping, the discretized form of these equations is given by Eqs.~\eqref{eq:wave_eq_disc}:
\begin{equation}
\begin{aligned}
    p_{i,j}(t+\delta t) &= (1 - k^p_{i,j}\delta t)\, p_{i,j}(t) 
    - c^2_{i,j}\delta t \left(\nabla \cdot \mathbf{o}\right)_{i,j}(t+\tfrac{\delta t}{2}), \\
    o_{x,i,j}(t+\delta t/2) &= \frac{1 - k^o_{i,j} \delta t/2}{1 + k^o_{i,j} \delta t/2} o_{x,i,j}(t - \delta t/2) 
    + \frac{\delta t}{\delta x (1 + k^o_{i,j} \delta t/2)} (p_{i,j}(t) - p_{i-1,j}(t)),
\end{aligned}
\label{eq:wave_eq_disc}
\end{equation}
where the indices $i$ and $j$ refer to spatial locations, and $(\nabla \cdot \mathbf{o})_{i,j}(t+\frac{\delta t}{2}) = \frac{o_{x,i+1,j}(t+\frac{\delta t}{2}) - o_{x,i,j}(t+\frac{\delta t}{2})}{\delta x} + \frac{o_{y,i,j+1}(t+\frac{\delta t}{2}) - o_{y,i,j}(t+\frac{\delta t}{2})}{\delta y}$ is the discretized divergence of $\mathbf{o}$ at the half timestep. A similar discretization applies for $o_{y}$. In the implementation, the small positive baseline pressure damping $k^p = 10^{-4}$ maintains weakly damped oscillations. The baseline velocity damping is set to $k^o = 0$ and is only altered when an explicit control manipulation is applied. The internal FDTD discretization uses a Courant-number parameter of $0.1$ to define the spatial steps $\delta x$ and $\delta y$, which keeps the numerical scheme safely within the stable regime. To ensure stability, the Courant number, which relates $\delta t$ to $\delta x$ and $\delta y$, must remain below 1.

This formulation also clarifies how oscillations arise in the model. Individual units are not assigned intrinsic sinusoidal dynamics or fixed beta-frequency oscillators. Instead, oscillatory activity emerges from the recurrent local coupling between the pressure-like state $p_{i,j}$ and the velocity-like states $o_{x,i,j}, o_{y,i,j}$: $p$ is updated from the local divergence of $\mathbf{o}$, while $\mathbf{o}$ is updated from spatial pressure differences. These alternating updates allow activity to propagate across the grid as damped traveling waves, and the repeated exchange between $p$ and $\mathbf{o}$ generates resonance modes whose timescales depend on the local propagation speed and damping. The model neurons therefore oscillate only as components of this coupled dynamical field, not as independent pacemakers.

\begin{figure*}[ht]
    \centering
    \includegraphics[width=\textwidth, trim={0cm 0 0cm 0cm}, clip]{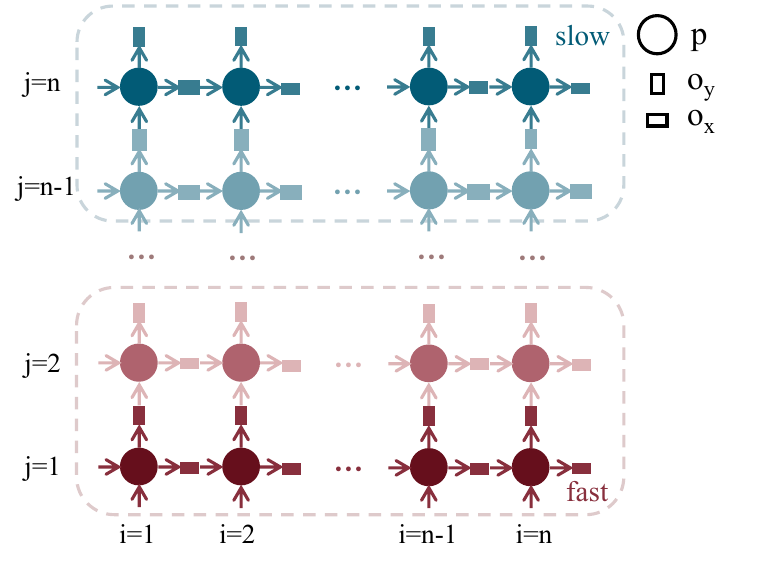}
    \caption{\textbf{Internal structure of each reservoir layer, implemented as a physics-based FDTD grid.} Each layer consists of primary neurons $p_{i,j}$ (circles) and intermediate neurons $o_{x,i,j}$ (horizontal rectangles) and $o_{y,i,j}$ (vertical rectangles), arranged on a staggered spatial grid with local sparse coupling (Eq.~\ref{eq:wave_eq}). The wave speed $c[i,j]$ varies along the $j$-axis according to $c[i,j] = c_0 + dc \cdot (i-1)$, creating a gradient from slow low-frequency dynamics (slow end, left) to fast high-frequency dynamics (fast end, right). The readout weights $W^{\text{tactus}}_{\text{out}}$ connect exclusively to the slow end of the grid, enabling the Tactus Layer to capture lower-frequency metric structure. Where present, input matrices are fixed, while the output weights $W^*_{\text{out}}$ are trained using MSE loss. The Tatum readout is pretrained first and then receives only an initial readout-level calibration for newly presented rhythms; during online calibration, layer-specific readouts are updated while the internal weights $W$ and applicable input matrices remain fixed across all rhythms.}
    \label{fig:2}
\end{figure*}

The two groups of unknowns can be interpreted as two types of artificial neurons within the reservoir, as illustrated in Fig.~\ref{fig:2}. The primary neuron is denoted as $p_{i,j}$, while the intermediate neuron is labeled as $o_{x,i,j}$ or $o_{y,i,j}$. These neurons can be grouped into a hidden state matrix, $\mathbf{x}$, as shown in Eq.~\eqref{eq:model}. Since $p$ and $\mathbf{o}$ are coupled locally and sparsely, the coupling matrix, $\mathbf{A}$, derived from Eq.~\eqref{eq:wave_eq_disc}, will also exhibit sparsity.

The weight matrix, $\mathbf{W}$, of the reservoir is computed as:
\begin{equation}
\mathbf{W} = \frac{\mathbf{A} - (1 - \alpha) \cdot \mathbf{I}}{\alpha},
\label{eq:W_matrix_calculation}
\end{equation}
where $\mathbf{I}$ is the identity matrix. In this formulation, the reservoir update Eqs.~\eqref{eq:W_matrix_calculation} closely resemble the FDTD update equations, implying strong symmetry constraints on the $\mathbf{W}$ matrix. The subtraction of \((1-\alpha)\mathbf{I}\) is required because \(\mathbf{W}\) is embedded inside the leaky reservoir update. In the locally linear regime, the effective transition matrix is \((1-\alpha)\mathbf{I}+\alpha\mathbf{W}\), which equals the FDTD update matrix \(\mathbf{A}\). Thus, even if some entries of \(\mathbf{W}\) appear strongly negative after rescaling by \(1/\alpha\), the effective state transition remains the intended FDTD operator. The local value of $c$ determines the responsiveness of the $p$-neuron to inputs from neighboring $o$-neurons. Together with the coupling to its neighbors, this can result in local resonances, where small $c$ values correspond to low-frequency resonances, a phenomenon akin to physical systems. In our implementation, each reservoir layer operates on a $40 \times 40$ spatial grid ($n = 40$), yielding $1{,}600$ primary neurons $p_{i,j}$ and $3{,}200$ intermediate neurons ($o_{x,i,j}$ and $o_{y,i,j}$), for a total of $4{,}800$ units per layer. The full hidden state vector $h_t$ in Eq.~\eqref{eq:model} therefore has dimension $3n^2 = 4{,}800$, and the structured weight matrix $W \in \mathbb{R}^{4800 \times 4800}$ is derived from the FDTD discretization in Eq.~\eqref{eq:wave_eq}.
Because $\mathbf{W}$ is derived from the FDTD transition matrix $\mathbf{A}$ through Eq.~\eqref{eq:W_matrix_calculation}, its diagonal entries remain close to 1. These near-unity values ensure that the effective transition $(1-\alpha)\mathbf{I}+\alpha\mathbf{W}$ reproduces $\mathbf{A}$.

By introducing a gradient in $c$ on top of random values, a reservoir with both slow (low-frequency resonances) and fast (high-frequency resonances) dynamics can be achieved, as shown in Fig.~\ref{fig:2}. Specifically, $c$ is initialized on an $n \times n$ 2D spatial grid combining a deterministic spatial gradient along the $i$-axis,
\begin{equation}
c[i,j] = c_0 + dc \cdot (i-1),
\end{equation}
where $c_0 = 300$~m/s and $dc = -250/n$~m/s per grid point (approximately $-6.25$~m/s for $n=40$), with a multiplicative random perturbation,
\begin{equation}
c[i,j] \leftarrow c[i,j] - c_{rr} \cdot \mathrm{rand}(c[i,j]),
\end{equation}
where $c_{rr} = 0.8$ and $\mathrm{rand}(\cdot) \sim \mathcal{U}(0,1)$ is sampled independently at each grid point. Here, $n=40$ defines a $40 \times 40$ grid per layer, giving $1{,}600$ pressure-like units and $3{,}200$ velocity-like units, for a total of $4{,}800$ states; this size was chosen as a compromise between dynamical richness and tractable computation. The baseline wave speed $c_0 = 300$ and gradient $dc=-250/n$ were selected to create a broad spatial ordering of intrinsic timescales, so that the same reservoir contains both faster tatum-like and slower tactus-/measure-like resonances. The perturbation strength $c_{rr}=0.8$ was chosen to break exact spatial symmetry and diversify the available modes without destroying the overall gradient. The deterministic gradient provides a structured frequency hierarchy, while the random perturbation (up to $80\%$ variation per location) breaks spatial symmetry and promotes diverse dynamical modes. The variable $k$ controls the information propagation speed between $p$-neurons via the $o$-neurons. Increasing $k$ results in more heavily damped resonances.

The model parameters were selected to satisfy dynamical and computational requirements. In particular, the grid size ($n=40$) was chosen to provide a sufficiently rich reservoir while remaining computationally tractable; the spatial wave-speed field ($c_0 = 300$~m/s with gradient $dc=-250/n$ and random perturbation $c_{rr}=0.8$) was chosen to produce a broad hierarchy of resonant timescales across the grid; and the damping and discretization parameters were set to keep the FDTD dynamics weakly damped and numerically stable. These parameters were not fitted to EEG recordings or tuned to a particular beta-band frequency. The beta-band structure reported in the Results is treated as a qualitative correspondence of the model dynamics.

\subsection*{Tatum Layer pretraining}

Before rhythm-specific calibration, the Tatum Layer is pretrained on synthetic pulse-train rhythms to obtain a rhythm-general initialization of $W_{\mathrm{out}}^{\mathrm{tatum}}$. Synthetic pretraining rhythms were generated following the artificial-rhythm construction principle used in previous dynamic-systems rhythm-interaction models~\cite{yuan2024novel,yuan2025dynamic}. Each sample was represented as a smooth pulse train. For isochronous samples, a single inter-beat interval was sampled from the predefined training range and kept fixed over the full sequence. For variable-IBI samples, successive intervals were randomly varied within the same range, producing non-isochronous pulse trains. This pretraining stage was used only to initialize the Tatum readout; the representative test rhythms analysed in the Results were not included as supervised training examples.

\subsection*{Online calibration protocol}
\label{sec:Online calibration protocol}

When the model is exposed to a new rhythm, the recurrent reservoir weights $W$ and input weights $W_{\mathrm{in}}$ remain fixed. Only the layer-specific readout weights $W_{\mathrm{out}}$ are adapted.

The simulation timestep is $\delta t = 6$ms, so the 200ms anticipatory horizon corresponds to 33 timesteps ($33 \times 6 \approx 198$ms). This anticipatory horizon is used for the Tatum and Motor prediction targets. The Tatum Layer is pretrained in a 200ms-ahead tatum-level reference frame and is then recalibrated at the readout level during the initial 0–6s window of a newly presented rhythm. The Motor Layer is trained and evaluated as a 200ms-ahead predictor of surface event onsets in the binary rhythm pattern.

The Tactus and Higher Cognition layers serve different functions and are not trained as 200~ms-ahead surface-event predictors. The Tactus Layer is calibrated to a tactus-like grouping target derived from the symbolic rhythm pattern. For a pattern of length $L$, all divisors of $L$ are considered as candidate grouping factors. Each candidate defines a periodic template of the form \texttt{1 0 … 0} repeated across the pattern cycle. The template with the highest alignment score with the surface-event positions is selected as the Tactus Layer calibration target. This procedure provides an algorithmically derived grouping target for adapting $W_{\mathrm{out}}^{\mathrm{tactus}}$ and does not pre-encode a manually annotated human tactus. For the ambiguous pattern \texttt{10010}, the selected best factor can be 1; in this case, the original pattern is used as the tactus-like target.

The Higher Cognition Layer is calibrated to predict event presence or absence at tatum-aligned positions. Its binary target is constructed directly from the repeating rhythm pattern: positions containing a surface event are assigned label 1, and silent positions are assigned label 0. The Higher Cognition readout is optimized using a binary cross-entropy loss. During inference, the resulting event/silence prediction is used as the top-down control signal for Motor Layer suppression at anticipated-but-silent tatum-level pulses.

After the calibration phase, all readout weights are frozen and the model proceeds in inference mode. Thus, rapid adaptation acts only on $W_{\mathrm{out}}$ and does not change the internal recurrent dynamics of the oscillatory reservoir.
\subsection*{Motor-layer output suppression via damping control}

Suppression of movement-timing predictions is implemented through transient modulation of the Motor Layer dynamics. During inference, all readout weights are fixed and no pulse-by-pulse weight updates occur. When the Higher Cognition Layer predicts that a tatum-aligned movement-timing prediction should be omitted, it sends a binary control signal that temporarily sets $k^p > 0$ to $k^p < 0$ in the Motor Layer. As established in the theoretical analysis, negative $k^p$ shifts the real part of the system eigenvalues to positive values, driving exponential divergence of the pressure field. This divergence is rapidly saturated by the tanh nonlinearity, disrupting the stable phase relationships required for coherent rhythmic output and thereby suppressing Motor Layer prediction pulses at predicted silent positions. Equivalently, the Motor Layer recurrent connectivity is switched from an oscillatory regime ($W^{\text{motor}}_{\text{ori}}$, corresponding to $k^p > 0$) to an unstable regime ($W^{\text{motor}}_{\text{damp}}$, corresponding to $k^p < 0$) for a fixed interval. After the suppression window, $k^p$ reverts to its original positive value, restoring stable oscillatory dynamics.

This intervention is restricted to the Motor Layer and does not affect the dynamics of the Tatum, Tactus, or Higher Cognition layers, which continue to encode temporal structure and predictions. In the state-space formulation, setting $k_p < 0$ destabilizes the Motor Layer so that oscillations no longer support coherent movement-timing prediction pulses. The resulting effect is suppression of the movement-timing prediction at the predicted silent tatum position. After the suppression window, the Motor Layer reverts to its original oscillatory regime.

\subsection*{Theoretical Framework}

The reservoir dynamics described in the Methods Section support rich oscillatory behavior, enabling the model to track and generate rhythmic patterns. The model also requires the ability to suppress movement-timing predictions at anticipated-but-silent tatum-level pulses or during syncopation. To provide this functionality, our model dynamically modulates the damping parameter \(k_p\), which governs the stability of internal oscillations. In this subsection, we present a theoretical analysis demonstrating how changes in \(k_p\) shift the system between oscillatory tracking and active inhibition modes, thereby providing a mathematically grounded mechanism for flexible rhythmic control.

\subsubsection*{Fourier Domain Representation}

To analyze the system behavior, we examine the wave equations in the frequency domain. For a time-dependent field \( f(t) \), the temporal and spatial derivatives become:
\begin{equation}
\frac{\partial f}{\partial t} \xrightarrow{\text{Fourier}} -i\omega \tilde{f}, \quad
\nabla f(\mathbf{r}) \xrightarrow{\text{Fourier}} i \mathbf{k} \tilde{f}
\end{equation}

where \( \mathbf{k} = [k_x, k_y] \) is the spatial wave vector and \( \omega \) is the angular frequency.

Substituting these into the wave system (Eq.~\eqref{eq:wave_eq}), the dynamics in Fourier space reduce to:
\begin{equation}
\begin{bmatrix}
-i\omega + k_p & c^2 i k_x & c^2 i k_y \\
i k_x & -i\omega + k_x & 0 \\
i k_y & 0 & -i\omega + k_y
\end{bmatrix}
\begin{bmatrix}
p \\ o_x \\ o_y
\end{bmatrix}
= 0
\end{equation}

\subsubsection*{Eigenvalue Structure and System Stability}

The characteristic frequencies \( \omega \) of the system are given by the solution of:
\begin{equation}
\det(\mathbf{A} - i\omega \mathbf{I}) = 0
\end{equation}

Expanding the determinant yields a cubic equation in \( \omega \). In the low-damping regime (small \( k_p, k_x, k_y \)), we obtain the simplified expression:
\begin{equation}
i \omega^3 - \left(k_p + k_x + k_y\right) \omega^2 - i \left(c^2 k_x^2 + c^2 k_y^2 + k_p k_x + k_p k_y + k_x k_y\right) \omega + c^2 k_x^2 k_y + c^2 k_x k_y^2 + k_p k_x k_y = 0
\label{eq:eigenvalue_combine}
\end{equation}

From this expression, two key observations follow:

\paragraph{(1) Damping Controls the Real Part:} The term \( -\left(k_p + k_x + k_y\right) \omega^2 \) shows that damping directly contributes to the real part of the eigenvalues. A positive sum results in decaying oscillations; a negative sum induces instability.

\paragraph{(2) Propagation Dominates the Imaginary Part:} The terms involving \( c^2 \) and \( |\mathbf{k}| \) dominate the imaginary components, resulting in oscillatory propagation with frequency approximately \( \pm c |\mathbf{k}| \).

\subsubsection*{Approximate Eigenvalues and Stability Conditions}

To leading order, the eigenvalues can be approximated by:
\begin{equation}
\omega \approx -\text{Re}(k_p) \pm i c |\mathbf{k}|
\label{eq:approx_eigenvalues}
\end{equation}

This reveals how the system's stability depends directly on the sign of \( k_p \):
\begin{align}
k_p > 0: &\quad \text{Damped oscillations (stable regime)}, \\
k_p = 0: &\quad \text{Marginally stable propagation}, \\
k_p < 0: &\quad \text{Exponential divergence (unstable regime)}.
\end{align}

When $k_p < 0$, the positive real part of the eigenvalues causes oscillations to grow unstably. In practice, the tanh nonlinearity rapidly saturates these diverging oscillations, disrupting the stable phase relationships required for coherent movement-timing prediction. The functional outcome is therefore suppression of the movement-timing prediction, even though the underlying dynamical mechanism is exponential divergence of the oscillatory state. After the suppression window, $k_p$ reverts to its positive value, restoring stable oscillatory dynamics.

\subsubsection*{Dynamic Control of Inhibition via \( k_p \)}

To leverage this mechanism for rhythmic inhibition, we set \( k_x = k_y = 0 \) so that all damping is controlled solely by \( k_p \). This simplifies the system's behavior and enables dynamic modulation of neural excitability through:
\begin{equation} \label{eq:kp_switch}
k_p = 
\begin{cases}
k_p^{+} & \text{(small positive value for maintaining stable oscillations)} \\
k_p^{-} & \text{(negative value for inducing active inhibition)}
\end{cases}
\end{equation}

When the Higher Cognition Layer predicts an upcoming anticipated-but-silent tatum-level pulse, the system transitions to an output-suppression regime by setting \( k_p = k_p^{-} \). This causes the real part of the eigenvalues to become positive, destabilizing the oscillatory state so that it no longer yields a coherent movement-timing prediction. In contrast, during regular tatum-level pulse prediction, \( k_p \) remains positive, preserving resonance and entrainment.

\subsubsection*{Functional Interpretation}

This modulation of \( k_p \) forms the theoretical foundation of our inhibition strategy. By exploiting the dynamic interplay between oscillatory propagation and damping-driven instability, the model flexibly suppresses motor layer activity at predicted silent intervals without compromising global rhythmic tracking.

This eigenvalue-based analysis highlights how tuning \(k_p\) modulates the real part of the system's eigenfrequencies, controlling the transition from stable rhythmic oscillations to exponential divergence. This mechanism enables the model to maintain entrainment during regular beats while suppressing movement-timing predictions at predicted silent positions. We use suppression of movement-timing predictions to describe this computational mechanism (\(k_p < 0\)); beta-band modulation is treated as a qualitative comparison with human rhythm studies. Overall, this framework provides the theoretical foundation for incorporating biologically inspired inhibitory control within a rhythm-sensitive reservoir system. Full derivations are available in the Supplementary Note 1.

All derivations and stability analyses are included in the Supplementary Note 1.

\subsection*{Layer-specific implementation details}

\begin{figure*}[ht]
    \centering
    \includegraphics[width=\textwidth, trim={0cm 0 0cm 0cm}, clip]{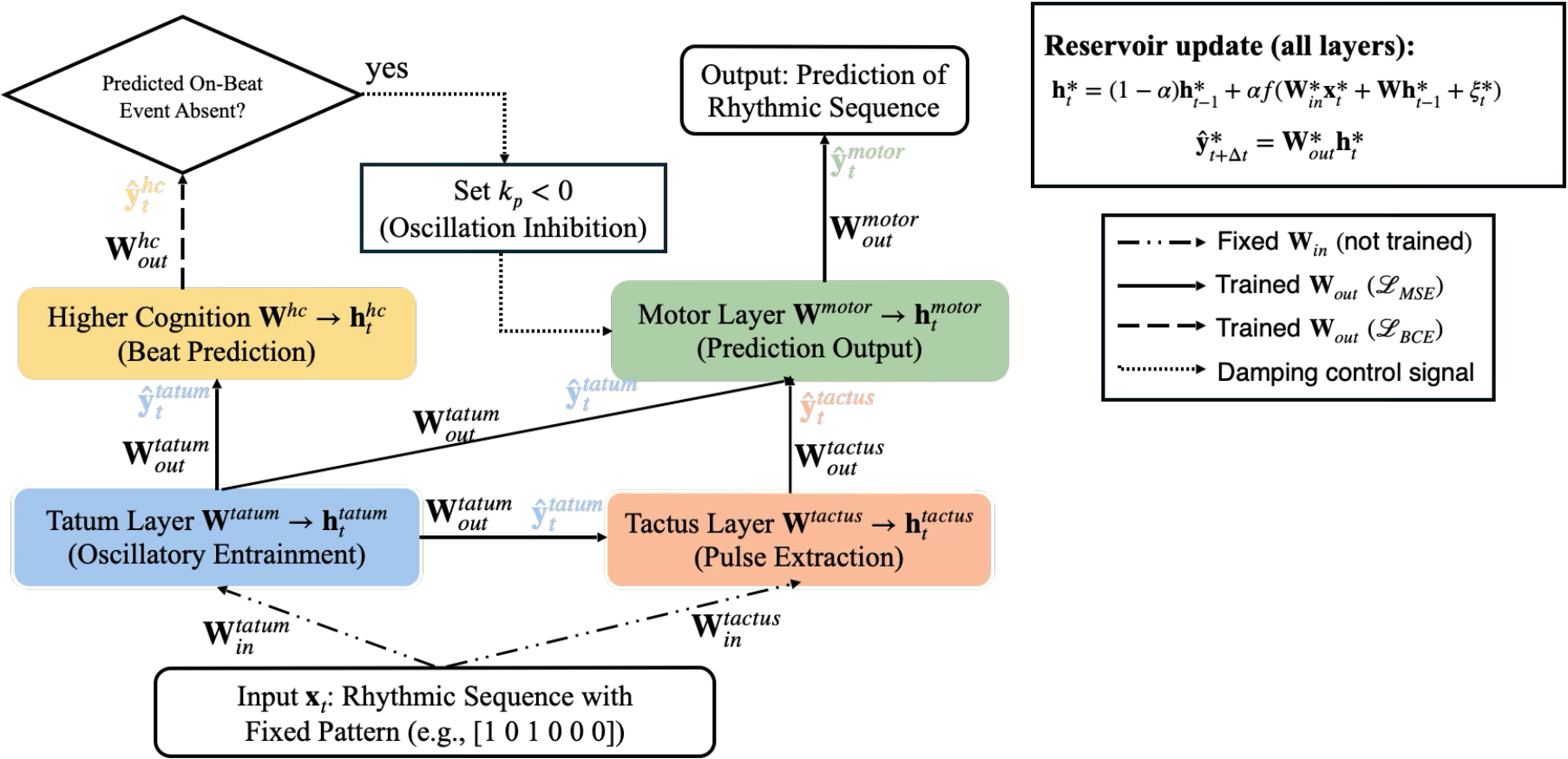}
    \caption{\textbf{Detailed information processing flow diagram illustrating the dynamic interactions between layers.} The Higher Cognition Layer predicts event presence or absence at tatum-aligned positions and modulates the Motor Layer damping coefficient ($k_p < 0$) to induce suppression of movement-timing predictions at predicted absent events. In this figure, the label ``Oscillation Inhibition'', where present within the embedded schematic, designates this movement-timing prediction suppression regime generated by destabilized oscillatory dynamics. The Tactus Layer receives both the raw rhythmic input and the Tatum Layer output $\hat{y}^{\text{tatum}}_t$, i.e., $\mathbf{x}^{\text{tactus}}_t = \hat{y}^{\text{tatum}}_t + \mathbf{x}^{\text{tatum}}_t$. The Motor Layer integrates $\mathbf{x}^{\text{motor}}_t = \hat{y}^{\text{tatum}}_t + \hat{y}^{\text{tactus}}_t$. All layers follow the reservoir update in Eq.~\ref{eq:model}. Where used, input matrices are fixed; $W^*_{\text{out}}$ are trained with $\mathcal{L}_{\text{MSE}}$ (solid arrows); $W^{\text{hc}}_{\text{out}}$ with $\mathcal{L}_{\text{BCE}}$; dotted arrows indicate damping control signals.}
    \label{fig:3}
\end{figure*}

The rhythmic pattern is not pre-encoded in the task-specific readouts, analogous to a novel rhythm being unfamiliar to a listener. The model therefore uses a short readout-level calibration period before inference. During this period, only $W_{\mathrm{out}}$ is updated, while $W$ and $W_{\mathrm{in}}$ remain fixed, as specified in the Online calibration protocol. The update procedure is implemented with a limited number of gradient-descent iterations per update step, as described in Algorithm~\ref{alg:learning_algorithm} and illustrated in Fig.~\ref{fig:3}.

Although $\hat{y}^{\text{tatum}}$, 
$\hat{y}^{\text{tactus}}$, and $\hat{y}^{\text{motor}}$ all contribute to the predictive hierarchy, their targets differ according to each layer's functional role. $\hat{y}^{\text{tatum}}$ is learned during pretraining and predicts the tatum-level pulse train in the 200~ms-ahead anticipatory reference frame. As detailed in the Online calibration protocol, the Tatum readout is then recalibrated for each new rhythm only during the initial 0--6 s window before the Tactus and Motor readouts are calibrated to their own functional targets: $\hat{y}^{\text{tactus}}$ is fitted to the algorithmically derived tactus-like grouping target from the symbolic pattern, and $\hat{y}^{\text{motor}}$ predicts the surface event onsets (the \texttt{1}s in the symbolic pattern). After this calibration window, the readouts are fixed and the Motor Layer generates the final anticipatory movement-timing prediction from the calibrated Tatum scaffold, the calibrated Tactus representation, and the Higher Cognition Layer's event-presence predictions.

Conceptually, these three targets should not be interpreted as equally difficult versions of the same task. The tatum target is intended to model a lower-level, rhythm-general temporal subdivision process that establishes the smallest shared time grid for the hierarchy. By contrast, the tactus-like target reflects a higher-level algorithmically derived grouping of tatum subdivisions, which is more pattern- and context-dependent and therefore benefits more directly from rhythm-specific readout adaptation. The Motor Layer then relies on this tatum-defined fine-grained scaffold and the calibrated tactus-like grouping to place anticipatory movement-timing predictions at event-level resolution.

The Tatum Layer is designed to capture the smallest rhythmic interval, or tatum, from the complex input pattern. Its readout is pretrained on a synthetic rhythm dataset using the tatum-level target in the shared 200~ms-ahead reference frame, and this pretrained readout then receives an initial 0--6 s output-weight calibration during online calibration for a new rhythm. As shown in Fig.~\ref{fig:2} and Fig.~\ref{fig:3}, the input to the Tatum Layer, $x_{t}^{tatum}$, represents the rhythm at time step $t$, with its output weights $W_{out}^{tatum}$ trained using MSE loss (solid lines in Fig.~\ref{fig:3}).

Human perception of rhythm typically requires slower processing, addressed by the Tactus Layer. This layer processes the Tatum Layer output in conjunction with the original complex pattern. The input to the Tactus Layer at time step $t$, $x_{t}^{tactus}$, is given by $\hat{y}^{tatum}_{t} + x_{t}^{tatum}$, as depicted in the information flow diagram of Fig.~\ref{fig:3}. As illustrated in Fig.~\ref{fig:2}, the reservoir structure incorporates fast and slow ends to capture fast and slow frequency envelopes, respectively. By inputting these signals into the fast end of the Tactus Layer and connecting the readout layer only to the slow end, the Tactus Layer provides a reservoir basis from which the output weights can be adapted to the algorithmically derived tactus-like grouping target. Its output weights $W_{out}^{tactus}$ are trained using MSE loss against this imposed grouping target, allowing the layer to express slower rhythmic groupings with reliable phase-locking.

A single-layer reservoir is insufficient to capture complex rhythmic structures due to its limited memory capacity. Therefore, we introduce the Higher Cognition Layer, which serves as a working memory, learning and storing rhythmic patterns. This layer is implemented as a state-space model and is activated by peaks in the Tatum Layer output, as shown in the trigger mechanism of Fig.~\ref{fig:3}. When a peak is detected, a trigger signal activates the Higher Cognition network, enabling it to predict whether the next tatum-aligned event onset will be present or absent. Peaks are identified when $\hat{y}^{tatum}_{t-1} \geq \hat{y}^{tatum}_{t-2}$ and $\hat{y}^{tatum}_{t-1} \geq \hat{y}^{tatum}_{t}$, with $\hat{y}^{tatum}_{t-1}$ marked as the event occurrence time. Upon receiving the trigger at position $l$, the hidden state $h_{l}^{hc}$ is updated, and the output weight $W_{out}^{hc}$ infers the probability of a missing or present event after one tactus- or pattern-relevant interval. During training, $W_{out}^{hc}$ is updated by minimizing the Binary Cross Entropy loss (dashed line in Fig.~\ref{fig:3}), with the learning of $W_{out}^{hc}$ beginning before the other layers engage in prediction. This process mirrors human participants learning a novel rhythm, requiring time to listen and comprehend the structure, thereby enhancing the anticipation of rhythmic variations, including the prediction of missing events.

Following the prediction from the Higher Cognition Layer, the neuron connections within the Motor Layer, denoted as $W^{motor}$, are adjusted through the dynamic damping control mechanism illustrated in Fig.~\ref{fig:3}. The Motor Layer integrates input from both the Tatum and Tactus Layers, where $x_{t}^{motor}$ is $\hat{y}^{tatum}_{t} + \hat{y}^{tactus}_{t}$. Its output weights are updated to align with the predictions, enabling the network to capture complex rhythmic structures and generate appropriate motor responses.

The Motor Layer may produce pulse-like movement-timing predictions at anticipated-but-silent tatum-level pulses, where the internal rhythmic structure creates a pulse-like expectation but the learned binary rhythm pattern specifies silence. The Higher Cognition Layer provides a top-down event/silence prediction for these positions. When a position is predicted to be silent, the Higher Cognition Layer modulates Motor Layer damping to suppress the inappropriate movement-timing prediction, as depicted in the inhibition control pathway of Fig.~\ref{fig:3}.

Damping is controlled by dynamically adjusting the connection matrix of the Motor Layer. When the Higher Cognition Layer predicts an anticipated-but-silent tatum-level pulse, the damping matrix $k$ is increased to a stronger negative value (set $k^p < 0$ in Fig.~\ref{fig:3}), generating the connectivity matrix $W^{motor\_damp}$. After a predefined period, the matrix reverts to its original state $W^{motor\_ori}$, allowing normal rhythmic behavior to resume.

After completing the calibration phase, the calibrated Tatum readout and the calibrated downstream output weights $W_{out}^{*}$ were fixed. At the designated prediction start time, the Higher Cognition Layer initiated the Motor Layer's connectivity state switching, as shown in the decision flow of Fig.~\ref{fig:3}. This matrix switching substantially modulated neuronal activity. We recorded neural responses across several common rhythmic patterns and compared these with human EEG spectra. The observed similarities between the neural activity produced by our model and that in human brain rhythms suggest that our network effectively replicates key aspects of human rhythmic perception.

\section*{Data Availability}
The datasets generated during the current study are available from the corresponding author. Alternatively, as the data used in this study were algorithmically generated using the methods described in the manuscript, the parameters and scripts used to produce these datasets will be provided in the code repository upon publication.

\section*{Code Availability}
The full source code for this study is publicly available in the following GitHub repository: \url{https://github.com/zjyuan1208/ESN_complex_rhythm_perception}.

\section*{Acknowledgements}

The authors Z.~Y., G.~W., and D.~B. are gratefully acknowledge the financial support for this work provided by the BOF grant (BOF/24J/2021/246) and the WithMe FWO grant (G0A0220N). This research was also partially supported by the Flemish AI Research Programme.

\section*{Author Contributions}

Conceptualization and methodology were performed by Z.~Y, D.~B, and G.~W.. And Z.~Y. developed the main algorithm and performed the experiments. Data curation and analysis were carried out by Z.~Y., D.~B, and G.~W.. The manuscript was drafted by Z.~Y. and revised critically for important intellectual content by D.~B, and G.~W. All authors reviewed the manuscript and approved the final submission.

\section*{Competing Interests}
The authors declare no competing financial or non-financial interests.

\bmhead{Supplementary information}

Supplementary Text\\
Eqs. S1 to S26\\

\bibliography{sn-bibliography}

\clearpage
\section*{Figures}

\clearpage
\section*{Tables}

\begin{table}[h!]
\centering
\scriptsize 
\caption{IBIs (ms) for Different Rhythm Patterns, Models, and BPMs for Tatum Output. This is used to replace the current ridar figure.}
\begin{tabular}{lcccccc}
\toprule
\multirow{2}{*}{Model} & \multicolumn{3}{c}{101001} & \multicolumn{3}{c}{101000} \\
\cmidrule(lr){2-4} \cmidrule(lr){5-7}
 & BPM 60 & BPM 100 & BPM 140 & BPM 60 & BPM 100 & BPM 140 \\
\midrule
GRU                & 266.88 & 184.99 & 152.95 & 383.10 & 228.81 & 181.11 \\
TFTransformer      & 271.66 & 235.30 & 158.22 & 586.76 & 223.10 & 205.22 \\
Ours (before)      & 568.90 & 544.67 & 322.69 & 615.94 & 458.91 & 527.25 \\
Ours (after)       & 497.31 & 900.00 & 421.50 & 750.57 & 600.00 & 420.27 \\
\midrule
\multirow{2}{*}{Model} & \multicolumn{3}{c}{10101001} & \multicolumn{3}{c}{1001001010} \\
\cmidrule(lr){2-4} \cmidrule(lr){5-7}
 & BPM 60 & BPM 100 & BPM 140 & BPM 60 & BPM 100 & BPM 140 \\
\midrule
GRU                & 264.48 & 174.37 & 148.96 & 316.73 & 193.22 & 150.41 \\
TFTransformer      & 262.63 & 209.62 & 150.02 & 427.42 & 189.65 & 151.36 \\
Ours (before)      & 604.71 & 456.74 & 307.95 & 699.64 & 514.21 & 330.27 \\
Ours (after)       & 497.81 & 482.33 & 419.89 & 497.61 & 428.78 & 525.41 \\
\midrule
\multirow{2}{*}{Model} & \multicolumn{3}{c}{10110010} & \multicolumn{3}{c}{101010100100} \\
\cmidrule(lr){2-4} \cmidrule(lr){5-7}
 & BPM 60 & BPM 100 & BPM 140 & BPM 60 & BPM 100 & BPM 140 \\
\midrule
GRU                & 262.37 & 170.11 & 141.01 & 299.10 & 184.76 & 148.13 \\
TFTransformer      & 275.97 & 191.46 & 157.67 & 455.40 & 186.68 & 154.74 \\
Ours (before)      & 568.49 & 407.42 & 330.77 & 701.93 & 483.00 & 298.88 \\
Ours (after)       & 498.13 & 498.13 & 420.00 & 748.57 & 515.39 & 503.21 \\
\bottomrule
\end{tabular}
\label{tab:combined_tatum_intervals}
\end{table}

\end{document}